\documentclass[aip,jcp,unsortedaddress,12pt,eqsecnum,reprint]{revtex4-1}

\usepackage{amsthm,amsfonts,amssymb,amsmath,dsfont}
\usepackage[T1]{fontenc}
\usepackage{graphicx}
\usepackage{graphics}
\usepackage{color}
\usepackage{colortbl}
\usepackage{natbib}
\usepackage[colorlinks=true,linkcolor=blue]{hyperref}

\begin{document}

\newcommand{\pd}[3]{\frac{\partial^{#3}{#1}}{\partial {#2}^{#3}}}
\newcommand{\pdmix}[3]{\frac{\partial^{2}{#1}}{\partial{#2}\partial{#3}}}
\newcommand{\od}[3]{\frac{\mathrm{d}^{#3}{#1}}{\mathrm{d} {#2}^{#3}}}
\newcommand{\bra}[1]{\langle #1 |}
\newcommand{\ket}[1]{| #1 \rangle}
\newcommand{\braket}[2]{\langle #1 | #2 \rangle}
\newcommand{\bran}[1]{\{ #1 |}
\newcommand{\ketn}[1]{| #1 \}}
\newcommand{\braketn}[2]{\{ #1 | #2 \}}
\newcommand{\dif}{\mathrm{d}}
\newcommand{\conj}{{*}}
\newcommand{\im}{i}
\newcommand{\ihbar}{\frac{\im}{\hbar}}
\newcommand{\smallfrac}[2]{{\textstyle\frac{#1}{#2}}}
\newcommand{\Schrodinger}{Schr\"{o}dinger~}

\title{Multiconfigurational quantum propagation with trajectory-guided
generalized coherent states}

\author{Adriano Grigolo}
\email{agrigolo@ifi.unicamp.br}
\affiliation{Instituto de F\'{i}sica Gleb Wataghin, Universidade Estadual de
Campinas, Brazil}

\author{Thiago F. Viscondi}
\email{viscondi@if.usp.br}
\affiliation{Instituto de F\'{i}sica, Universidade de S\~{a}o Paulo, Brazil}

\author{Marcus A. M. de Aguiar}
\email{aguiar@ifi.unicamp.br}
\affiliation{Instituto de F\'{i}sica Gleb Wataghin, Universidade Estadual de
Campinas, Brazil}

\date{\today}

\begin{abstract}
A generalized version of the coupled coherent states method for coherent states
of arbitrary Lie groups is developed. In contrast to the original formulation,
which is restricted to frozen-Gaussian basis sets, the extended method is
suitable for propagating quantum states of systems featuring diversified
physical properties, such as spin degrees of freedom or particle
indistinguishability. The approach is illustrated with simple models for
interacting bosons trapped in double- and triple-well potentials, most
adequately described in terms of $\mathrm{SU}(2)$ and $\mathrm{SU}(3)$ bosonic
coherent states, respectively.
\end{abstract}



\maketitle

\section{Introduction}
\label{sec:introduction}

A vast number of physical systems exhibit the property that some of their parts
behave in a sort of classical way, meaning that quantum effects play only
a minor role in the description of those parts. This distinctive classical
character of specific degrees of freedom is a much welcomed attribute, for it
makes possible the development of tractable computational approaches capable of
carrying out the time-evolution of complex quantum systems, being thus the
fundamental property upon which time-dependent trajectory-guided methods are
based.

In this kind of technique quantum states are represented in terms of
time-dependent basis functions or `configurations'. Within a single
configuration, those degrees of freedom in which quantum effects are negligible
are evolved according to classical equations of motion. This classical dynamics
may be prescribed in a number of different ways and different choices correspond
to different propagation schemes.

In spite of the fact that individual configurations have some of their parts
bound to obey classical laws, a complete quantum solution is in principle
attainable by combining many configurations. The key idea behind such
`multiconfigurational' approaches is that trajectory-guided basis functions are
more likely to remain in the important regions of the Hilbert space, thus being
more efficient at representing the quantum state in the sense that a reduced
number of basis elements is required in order to achieve an accurate
description. And it is precisely through a significant reduction in the number
of basis functions needed to propagate the system that one hopes to escape the
exponential scaling of basis-set size with dimensionality typical of standard
static-basis formulations. This `mixed quantum-classical' picture is adopted in
many methods of quantum chemistry. \cite{Makri1999}

A recurrent theme in this field is the development of techniques which, by means
of equally simple recipes to guide the basis functions, would be readily
applicable to systems presenting authentically non-classical qualities, such as
spin degrees of freedom or particle exchange symmetry. Several works have been
directed to that purpose, most often aiming at a time-dependent description of
the electronic structure of molecules during non-adiabatic processes. One
particular example of such a recipe is the classical model for electronic
degrees of freedom proposed by Miller and White \cite{Miller1986} in which a
second-quantized fermionic Hamiltonian is properly reduced to a classical
function wherein number and phase variables play the role of generalized
coordinates. In contrast, a more `mechanistic' approach to fermion dynamics is
found on the multiconfigurational formula proposed by Kirrander and Shalashilin
\cite{Kirrander2011} in which the basis functions consist of antisymmetrized
frozen Gaussians \cite{Note01} guided by fermionic molecular dynamics.
\cite{Feldmeier2000}

Yet if one seeks to describe non-classical degrees of freedom by means of
classical-like variables, then generalized coherent states -- defined in the
group-theoretical sense -- are indisputably the most appropriate tools to be
employed. There are many reasons supporting this assertion.

First of all, coherent states are defined in terms of non-redundant parameters
and equations of motion for these parameters can be readily obtained from the
time-dependent variational principle. \cite{Kramer} In this way an optimized
time-evolution can be assigned in an unambiguous manner. Moreover, they are
naturally able to capture the desired symmetries of the system which are
maintained during propagation. Furthermore, the coherent-state parameters evolve
in a classical phase space in the strict sense of the word, hence we
automatically have at our disposal the wealth of analytical techniques
applicable to Hamiltonian systems. At the same time, through this intimate
connection to classical dynamics, coherent states provide a compelling classical
interpretation to quantum phenomena, in so far as individual configurations are
chosen to represent familiar objects -- i.e.~in such a way that it is meaningful
to discuss the dynamics of the system in terms of their trajectories. To this
extent, coherent states -- which are also minimum uncertainty states
(provided a proper meaning is assigned to the term `uncertainty')
\cite{Delbourgo77a, Delbourgo77b} -- are valuable tools in enhancing our
comprehension with respect to the semiclassical features of the quantum system
under investigation. In addition, the group-theoretical formalism secures a
well-defined integral form for the coherent-state closure relation
\cite{Zhang1990a} -- a crucial element to the developments hereby presented.
This list of virtues is not exhausted and other advantages of a generalized
coherent-state representation will be pointed out throughout the paper.

Along these lines, Van Voorhis and Reichman \cite{VanVoorhis2004} have
considered a number of alternative representations of electronic structure
making use of different coherent-state parametrizations and also examined their
adequacy to a variety of systems. \cite{Note02} Within the context of
non-adiabatic molecular dynamics, a particularly interesting coherent-state
representation \cite{Thouless1960, Deumens1989a} is found in the simplest and
most throughly investigated version of the Electron-Nuclear Dynamics theory,
developed by Deumens, {\"{O}}hrn and collaborators. \cite{Deumens1992,
Deumens1994} The same kind of coherent state has been discussed in detail,
within the field of nuclear physics, by Suzuki and Kuratsuji. \cite{Suzuki1983,
Kuratsuji1980a, Kuratsuji1983} Turning to bosonic dynamics, a semiclassical
trajectory-based formula in the $\mathrm{SU}(n)$ coherent-state representation
has been recently derived and successfully applied to a model of trapped bosons.
\cite{Viscondi2011, Viscondi2011b}

These methods are representative of the kind of technique one has in mind when a
description of intrinsically quantum degrees of freedom in terms of
classical-like variables is desired. However, they either constitute approximate
single-configuration approaches \cite{Deumens1992, Deumens1994} or involve
complicated trajectories that live in a duplicated phase space,
\cite{Viscondi2011, Viscondi2011b} sometimes relying on sophisticated
root-search techniques in order to determine them. \cite{VanVoorhis2004,
VanVoorhis2002, VanVoorhis2003} It seems that a multiconfigurational,
generalized coherent-state approach, based on simple -- as opposed to duplicated
-- phase-space trajectories would be more in the spirit of the familiar
time-dependent guided-basis methods of quantum chemistry. \cite{Note03} This is
precisely the direction we take here.

In this work a quantum initial value representation method which employs a
generalized coherent-state basis set guided by classical trajectories is
formulated. The resulting propagation scheme is regarded as a natural extension
of the coupled coherent states technique of Shalashilin and Child
\cite{Shalashilin2000, Shalashilin2001b, Shalashilin2004a} in so far as (i)
basis-set elements represent localized quantum states; (ii) each element
evolves independently in a generalized classical phase space and carries an
action phase; and (iii) the quantum amplitudes associated with individual
elements obey fully coupled equations of motion which present a number of
attractive qualities.

We begin in \S\ref{sec:cs-formalism} with an overview on generalized coherent
states, deliberately avoiding the underlying group-theoretical formalism
associated with their construction. In particular, we demonstrate how the
time-dependent variational principle leads to classical equations of motion for
the coherent-state parameters in a curved phase space. This preliminary
discussion is illustrated with specific examples. Next, in
\S\ref{sec:ccs-generalized}, we set forth to derive the working equations of the
generalized coupled coherent states method, first in continuum form and then in
terms of a discrete basis set. In the latter case, both unitary and non-unitary
versions of the formulas are devised. A primitive method, resulting from
propagation of a single configuration, is also discussed. In
\S\ref{sec:ccs-application} we use our approach to study model Hamiltonians and
compare our results against exact quantum data. We also take the opportunity to
expose certain particular aspects of the generalized coherent-state formulation.
Finally, conclusions are drawn in \S\ref{sec:conclusion}.

\section{Generalized coherent-state formalism}
\label{sec:cs-formalism}

Coherent states are most elegantly discussed within the context of group theory
and this is the point of view adopted here. We shall not venture into the
group-theoretical formalism itself though -- on that subject the reader is
referred to the instructive review by Zhang, Feng and Gilmore \cite{Zhang1990a}
or to a recent paper by the present authors. \cite{Viscondi2015} Here, we will
rather follow a more pragmatic approach according to which a coherent state is
given in the form of an expansion in a proper set of orthonormal states and work
its geometrical properties thereon.

\subsection{Preliminaries}
\label{sub:preliminaries}

Coherent states are Hilbert space vectors labeled by a complex vector $z = (z_1,
\ldots , z_d)$. \cite{Note04} They can be understood as the result of a
Lie-group operator acting on a reference state, which is recovered by setting
all entries of the vector $z$ to zero. We shall denote a non-normalized coherent
state by $\ketn{z}$. These curly \textit{ket} states are analytical in $z$,
while the \textit{bra} states $\bran{z}$ are analytical in the complex conjugate
variable, denoted by $z^\conj$; the normalized state $\ket{z}$ is not analytical
in $z$ for it depends on $z^\conj$ through the normalization factor
${\braketn{z}{z}}^{-\frac{1}{2}}$. The length $d$ of $z$ will be identified as
the number of degrees of freedom of the classical phase space associated with
the dynamics of $z$.

Coherent states of different groups are characterized by their distinct
geometrical properties which, in turn, are described in terms of a function $f$
related to the scalar product between two non-normalized coherent states:
\begin{equation}\label{eqn:f-def}
	f(z^\conj, z')
	= \log \braketn{z}{z'}.
\end{equation}
The \textit{classical phase-space metric} $g_{\alpha\beta}$ is a hermitian
matrix obtained by taking the cross derivatives of the real function $f(z^\conj,
z)$ with respect to its complex arguments, treating $z$ and $z^\conj$ as
independent variables: \cite{Note05}
\begin{equation}\label{eqn:g-def}
	g_{\alpha\beta}(z^\conj, z)
	=\pdmix{f(z^\conj, z)}{z_\alpha}{z^\conj_\beta}.
\end{equation}
The (non-orthogonal) coherent states span an overcomplete basis of the
corresponding Hilbert space and a closure relation holds:
\begin{equation}\label{eqn:closure}
	\hat{1}
	=\int \dif\mu(z^\conj,z) \ket{z}\bra{z},
\end{equation}
where the integration domain depends on the specific type of coherent state
being considered -- for semisimple compact Lie Groups or the Heisenberg-Weyl
group, for example, it extends over the entire $d$-dimensional complex plane. In
\eqref{eqn:closure}, the \textit{integration measure} $\dif\mu(z^\conj,z)$ is
defined as below:
\begin{equation}\label{eqn:measure}
	\dif \mu(z^\conj, z)
	=\kappa \det[g(z^\conj, z)] \prod^{d}_{\alpha=1} \frac{\dif^2 z_\alpha}{\pi},
\end{equation}
where $\dif^2 z_\alpha = \dif(\mathrm{Re}\: z_\alpha) \dif(\mathrm{Im}\:
z_\alpha) $ and the constant $\kappa$ is determined by normalization of
\eqref{eqn:closure} -- e.g.~by setting the expectation value of
\eqref{eqn:closure} in the reference state to unity -- and therefore it depends
on the quantum numbers that characterize the particular Hilbert space that
carries the coherent-state representation (some examples are found below in
\S\ref{sub:cs-examples}). In order to shorten the notation, we shall write
simply $\dif\mu(z)$, but keeping in mind that the measure is a real function of
both $z$ and $z^\conj$.

\subsection{Classical dynamics}
\label{sub:cl-dynamics}

The norm-invariant Lagrangian \cite{Kramer} that gives rise to the \Schrodinger
equation is
\begin{equation}\label{eqn:lagrangian-0}
	L(\psi)
	=\frac{\im\hbar}{2}\left[ \frac
	{\braket{\psi}{\dot{\psi}} -\braket{\dot{\psi}}{\psi}}
	{\braket{\psi}{\psi}} \right]
	-\frac{\bra{\psi} \hat{H} \ket{\psi}}{\braket{\psi}{\psi}},
\end{equation}
where $\ket{\psi} = \ket{\psi(t)}$ is an arbitrary quantum state and $\hat{H}$
represents the system's Hamiltonian operator. The time-dependent variational
principle (TDVP) states that, given a parametrization of $\ket{\psi}$ in terms
of some set of variables, the Euler-Lagrange equations obtained from
\eqref{eqn:lagrangian-0} translate into equations of motion for those same
variables. If the parametrization is flexible enough an exact quantum solution
is achieved. On the other hand, if the parametrization contains less than the
number of variables needed to span the associated Hilbert space, the resulting
dynamics will be only approximate.

We shall \textit{define} the \textit{classical equations of motion} for $z$ as
those equations obtained from the TDVP when a trial state $\ket{\psi} =
\ketn{z}$ is substituted in \eqref{eqn:lagrangian-0} -- that is, a trial state
whose dynamics is restricted to the nonlinear subspace consisting only of
coherent states. In this case, equation \eqref{eqn:lagrangian-0} takes the form:
\begin{equation}\label{eqn:lagrangian-1}
	L(z)
	=\frac{\im\hbar}{2} \sum^d_{\alpha=1} \left[
	\pd{f(z^\conj, z)}{z_\alpha}{} \dot{z}_\alpha
	-\pd{f(z^\conj, z)}{z^\conj_\alpha}{} \dot{z}^\conj_\alpha \right]
	-H(z^\conj, z),
\end{equation}
where the \textit{classical Hamiltonian} is the diagonal element of the operator
$\hat{H}$ in the coherent-state representation:
\begin{equation}\label{eqn:hamiltonian}
	H(z^\conj, z)
	=\frac{\bran{z} \hat{H} \ketn{z}}{\braketn{z}{z}}
	=\bra{z} \hat{H} \ket{z}.
\end{equation}

By means of the Euler-Lagrange equations one immediately finds that the dynamics
of $z$ obeys:
\begin{equation}\label{eqn:z-dot-1}
	\sum^d_{\beta=1} \dot{z}_{\beta} g_{\beta\alpha}(z^\conj, z)
	=-\ihbar \pd{H(z^\conj, z)}{z^\conj_\alpha}{},
\end{equation}
and that an equivalent (complex conjugate) equation holds for $z^\conj$. Notice
how the coherent-state geometry introduces a curvature in phase space by means
of the metric $g$. One now can distinguish between two kinds of coupling between
the components of the vector $z$: a \textit{dynamical coupling} via $H$, and a
\textit{geometrical coupling} induced by $g$.

The group-theoretical formalism assures us that Eq. (\ref{eqn:z-dot-1})
describes a Hamiltonian system in the most strict sense: the phase space
exhibits a symplectic structure, since a nondegenerate Poisson bracket can
always be established. \cite{Zhang1990a}

Furthermore, the measure \eqref{eqn:measure} is invariant under the classical
dynamics given by equation \eqref{eqn:z-dot-1}, that is, $\dif \mu(z(t_2)) =
\dif \mu(z(t_1))$, for any two instants $t_1$ and $t_2$ -- a property that we
recognize as a generalized form of the Liouville theorem and that remains valid
even when the system's Hamiltonian has an explicit time dependence.
\cite{Viscondi2013}

Finally, we define the \textit{complex action} $A$:
\begin{multline}\label{eqn:action-1}
	A(z^\conj(t), z(0), t)
	\\
	= S(z)
	-\smallfrac{\im\hbar}{2} \left[
	f(z^\conj(t), z(t)) +f(z^\conj(0), z(0)) \right],
\end{multline}
where the first term is the time integral:
\begin{equation}\label{eqn:action-0}
	S(z)
	= \int^t_0 L(z) \:\dif \tau,
\end{equation}
with the Lagrangian $L(z)$, given by equation \eqref{eqn:lagrangian-1}, being
evaluated over a classical orbit, satisfying \eqref{eqn:z-dot-1}.

As can be seen in \eqref{eqn:action-1}, the complex action $A$ carries imaginary
surface terms, which ensure that this is a well-defined analytical function on
both its complex arguments: $z^\conj(t)$ and $z(0)$, and also the time $t$. The
derivatives with respect to each of these variables are:
\begin{subequations}\label{eqn:action-derivatives}
\begin{align}
	\ihbar \pd{A(z^\conj(t), z(0), t)}{z^\conj_\alpha(t)}{}
	&=\pd{f(z^\conj(t), z(t))}{z^\conj_\alpha(t)}{},
	\label{eqn:action-derivatives-a}
	\\
	\ihbar \pd{A(z^\conj(t), z(0), t)}{z_\alpha(0)}{}
	&=\pd{f(z^\conj(0), z(0))}{z_\alpha(0)}{},
	\label{eqn:action-derivatives-b}
	\\
	\pd{A(z^\conj(t), z(0), t)}{t}{}
	&=-H(z^\conj(t), z(t)).
	\label{eqn:action-derivatives-c}
\end{align}
\end{subequations}
The above relations are recognized as the signature of a properly defined
classical action integral. \cite{Viscondi2015} Yet it is in terms of the real
quantity $S$ of Eq. \eqref{eqn:action-0} that our results are most conveniently
expressed. \cite{Note06} Therefore we shall denominate $S$ the \textit{action}.

\subsection{Examples of coherent states}
\label{sub:cs-examples}

In order to illustrate the formalism presented above, we consider simple
examples of coherent states and evaluate some of their geometrical elements,
such as the metric matrix $g$ and integration measure $\dif\mu$.

\subsubsection{Canonical coherent states}
\label{subsub:canonical-cs}

Canonical coherent states have their functional definition given in terms of a
superposition of bosonic Fock states with \textit{unrestricted} occupation
numbers; if the Hilbert space comprises $n$ modes, then the non-normalized
canonical coherent state is defined by:
\begin{equation}\label{eqn:weyl-cs-1}
	\ketn{z}
	=\sum^{\infty}_{m_1=0} \ldots \sum^{\infty}_{m_n=0}
	\left[ \prod^{n}_{\alpha=1}
	\frac{z^{m_\alpha}_\alpha}{\sqrt{{m_\alpha}!}} \right]
	\ket{m_1, \ldots, m_n},
\end{equation}
where the vacuum $\ket{0, \ldots, 0}$ is the reference state and the
length of the vector $z$ equals the number of bosonic modes: $d=n$.

Since the modes are assumed to be orthonormal, it follows immediately from
\eqref{eqn:weyl-cs-1} that the overlap between canonical coherent states is
\begin{equation}\label{eqn:weyl-overlap}
	\braketn{z}{z'}
	=\exp\left( \sum^n_{\alpha=1} z^\conj_\alpha z'_\alpha \right);
\end{equation}
thus, using \eqref{eqn:f-def}, we identify the function $f$ as:
\begin{equation}\label{eqn:weyl-f}
	f(z^\conj, z')
	= \sum^n_{\alpha=1} z^\conj_\alpha z'_\alpha.
\end{equation}
From \eqref{eqn:g-def}, the phase-space metric matrix $g$ is simply the identity
matrix:
\begin{equation}\label{eqn:weyl-g}
	g_{\alpha\beta}(z^\conj, z)
	=\delta_{\alpha\beta},
\end{equation}
which means that canonical coherent states give rise to a flat phase space and
therefore the degrees of freedom are not geometrically coupled. The normalized
measure, as defined in \eqref{eqn:measure}, is then trivial:
\begin{equation}\label{eqn:weyl-measure}
	\dif\mu(z)
	= \prod^n_{\alpha=1} \frac{\dif^2 z_\alpha}{\pi}.
\end{equation}

In what concerns semiclassical trajectory-based methods, canonical coherent
states are undoubtedly the most widely used type of coherent state. This is so
because of the following well-known homomorphism connecting the ladder operators
$(a^\dagger_\alpha, a_\alpha)$ of each bosonic mode in \eqref{eqn:weyl-cs-1}
with the position and momentum operators $(\hat{q}_\alpha, \hat{p}_\alpha)$:
\begin{equation}
	\hat{q}_\alpha
	=\frac{\gamma_\alpha}{\sqrt{2}}(a^\dagger_\alpha +a_\alpha),
	\ \ %
	\hat{p}_\alpha
	=\frac{\im\hbar}{\gamma_\alpha\sqrt{2}}(a^\dagger_\alpha -a_\alpha),
\end{equation}
with $[a_\alpha, a^\dagger_\beta] = \delta_{\alpha\beta}$ and where
$\gamma_\alpha$ is an arbitrary constant that sets the appropriate length scale
in each mode. The position representation of $\ket{z}$ is then found to be a
multidimensional Gaussian wavepacket, whose mean position $q$ and mean momentum
$p$ are related to the real and imaginary parts of the complex vector $z$,
respectively. Furthermore, the dynamics of $(q,p)$, as obtained from the
time-dependent variational principle, is simply given by Hamilton's classical
equations of motion in canonical form, the Hamiltonian being the mean value
given by equation \eqref{eqn:hamiltonian}. This so-called `frozen-Gaussian
representation' provides an obviously suitable framework for semiclassical
applications. \cite{Heller1981b, Heller1991, Herman1984, Kluk1986, Herman1994,
Kay1994, Baranger01, Kay2006, Aguiar2010b}

\subsubsection{\texorpdfstring{$\mathrm{SU}(n)$}{SU(n)} bosonic coherent states}
\label{subsub:sun-cs}

The $\mathrm{SU}(n)$ bosonic coherent states are suitable for describing systems
in a Fock space comprising $n$ modes and a \textit{fixed} total particle number
$N$. Their non-normalized form is
\begin{equation}\label{eqn:sun-cs-1}
	\ketn{z}
	={\sum_{\{m\}}}'
	\left( \frac{N!}{m_1! \ldots m_n!} \right)^{\frac{1}{2}}
	\left[ \prod^{n-1}_{\alpha=1} z^{m_\alpha}_\alpha \right]
	\ket{m_1, \ldots, m_n},
\end{equation}
where the reference state is $\ket{0, \ldots, N}$. In \eqref{eqn:sun-cs-1}, the
primed summation symbol means that the set of occupation numbers $\{m\}$ must
satisfy the condition $m_1 + m_2 + \ldots +m_n = N$. Because of this constraint,
the number of entries of the vector $z$ is one less than the number of modes:
$d=n-1$.

The overlap is easily evaluated from \eqref{eqn:sun-cs-1} with the help of the
multinomial theorem:
\begin{equation}\label{eqn:sun-overlap}
	\braketn{z}{z'}
	=\left[ 1 + \sum^{n-1}_{\alpha=1} z^\conj_\alpha z'_\alpha \right]^N;
\end{equation}
hence, by means of \eqref{eqn:f-def}, all geometrical aspects of these coherent
states are codified in the function $f$ given by:
\begin{equation}\label{eqn:sun-f}
	f(z^\conj, z')
	=N \log\left[1 + \sum^{n-1}_{\alpha=1} z^\conj_\alpha z'_\alpha \right].
\end{equation}

The metric matrix, according to \eqref{eqn:g-def}, is:
\begin{equation}\label{eqn:sun-g}
	g_{\alpha\beta}(z^\conj, z)
	=N \frac
	{(1 +|z|^2) \delta_{\alpha\beta} -z^\conj_\alpha z_\beta}{(1 +|z|^2)^2},
\end{equation}
where $|z|^2 = \sum^{n-1}_{\gamma=1} z^\conj_\gamma z_\gamma$. Clearly, the
fixed particle number condition translates into a geometrical coupling among the
components of the vector $z$.

Despite the complications introduced by the curved geometry, the metric's
determinant can be evaluated and the integration measure, defined in
\eqref{eqn:measure}, is found to be:
\begin{equation}\label{eqn:sun-measure}
	\dif\mu(z)
	=\frac{(N +n -1)!}{N! (1 +|z|^2)^n}
	\prod^{n-1}_{\alpha=1} \frac{\dif^2 z_\alpha}{\pi}.
\end{equation}

Recently, semiclassical methods employing $\mathrm{SU}(n)$ bosonic coherent
states, including an initial value representation based on classical
trajectories in a duplicated phase space, have been developed and tested with
an SU(3) model Hamiltonian \cite{Viscondi2011, Viscondi2011b} -- in
\S\ref{sub:su3-triple-well} we shall have the opportunity to revisit that same
problem.

\subsubsection{Spin coherent states}
\label{subsub:su2-cs}

A particularly interesting $\mathrm{SU}(n)$ coherent state originates when the
bosonic Fock space has only $n = 2$ modes. The $(N+1)$ states
\begin{equation*}\label{eqn:su2-basis}
	\ket{N,0},\ket{N-1,1},\ket{N-2,2},\ldots,\ket{1,N-1},\ket{0,N}
\end{equation*}
can be put into one-to-one correspondence with the well-known simultaneous
eigenstates $\ket{J, M}$ of the angular momentum operators $\{\hat{J}^2,
\hat{J}_z\}$ for a fixed total angular momentum $J = N/2$. In this way, the
non-normalized $\mathrm{SU}(2)$ coherent states \cite{Arecchi72,Zhang1990a} can
be expressed as:
\begin{equation}\label{eqn:su2-cs-1}
	\ketn{z}
	=\sum^{J}_{M= -J}{\binom{2 J}{J +M}}^{\frac{1}{2}} z^{J +M} \ket{J, M},
\end{equation}
which are especially designated as \textit{atomic} or \textit{spin coherent
states}. Notice that, in this specific case, the complex vector label $z$ has
dimension $d=1$ and $\ket{J, -J}$ is the reference state.

The overlap $\braketn{z}{z'}$ is easily evaluated using the orthogonality of the
$\ket{J, M}$ states. Alternatively, we can simply set $n = 2$ and $N = 2J$ in
equation \eqref{eqn:sun-overlap}, thus obtaining:
\begin{equation}\label{eqn:su2-overlap}
	\braketn{z}{z'}
	=(1 +z^\conj z')^{2 J},
\end{equation}
which leads to
\begin{equation}\label{eqn:su2-f}
	f(z^\conj, z')
	=2 J \log(1 +z^\conj z').
\end{equation}

According to equation \eqref{eqn:g-def}, it follows that the phase-space metric
$g$ (in this case, a scalar) is simply:
\begin{equation}\label{eqn:su2-g}
	g(z^\conj, z)
	=\frac{2J}{(1+|z|^2)^2}.
\end{equation}
The normalized measure is then found to be
\begin{equation}\label{eqn:su2-measure}
	\dif\mu(z)
	=\frac{(2J + 1)}{(1+ |z|^2)^2}
	\frac{\dif^2 z}{\pi}.
\end{equation}

The natural topology of the spin coherent state is that of the surface of a
sphere. In practical applications, one typically writes $z$ in terms of angles
$\theta$ and $\phi$:
\begin{equation}\label{eqn:su2-theta-phi}
	z
	=\tan(\theta/2) e^{-\im\phi},
\end{equation}
where $\theta \in [0, \pi]$ and $\phi \in [0, 2\pi)$. In these coordinates the
integration measure \eqref{eqn:su2-measure} reads:
\begin{equation}\label{eqn:su2-measure-theta-phi}
	\dif\mu(\theta, \phi)
	=(2J +1) \:\sin\theta\: \frac{\dif\theta \dif\phi}{4\pi}.
\end{equation}

Spin coherent states are discussed in more detail in \S\ref{sub:su2-double-well}
where we investigate a test model consisting of an $\mathrm{SU}(2)$ Hamiltonian.

\subsubsection{\texorpdfstring{$\mathrm{SU}(n)$}{SU(n)} fermionic coherent
states}
\label{subsub:thouless-cs}

Fermionic coherent states of the special unitary group are suitable for
describing a number-conserving system of $N$ fermions which are allowed to
occupy a set of $n$ orthonormal single-particle states ($n > N$). While
ultimately arbitrary, these underlying single-particle states are often taken to
be eigenstates of the non-interacting part of the Hamiltonian  or a set of
Hartree-Fock spin orbitals. \cite{Zhang1990a, Deumens1989a} A reference state
$\ket{\Phi_0}$ is specified by constructing a Slater determinant out of $N$ of
such spin orbitals (e.g.~the ones having lowest energies). These are denoted by
$\ket{\phi^\bullet_\alpha}$ with $1\leq\alpha\leq N$, and are said to belong to
the \textit{occupied} space. The remaining $M \equiv n - N$ spin orbitals,
denoted by $\ket{\phi^\circ_\mu}$ with $1\leq\mu\leq M$, are said to belong to
the \textit{virtual} space. Then, the non-normalized fermionic coherent state
can written as:
\begin{equation}\label{eqn:thouless-cs-1}
	\ketn{z} =
	\hat{A}_N \prod^{N}_{\alpha=1} \left[
	\ket{\phi^\bullet_\alpha}
	+\sum^{M}_{\mu=1} \ket{\phi^\circ_\mu} z_{\mu\alpha} \right],
\end{equation}
where the symbol $\hat{A}_N$ instructs anti-symmetrization of the direct product
of the $N$ single-particle states ${\ket{\phi^\bullet_\alpha} +\sum^{M}_{\mu=1}
\ket{\phi^\circ_\mu} z_{\mu\alpha}}$, which are sometimes called
\textit{dynamical orbitals}. \cite{Deumens1994}

In the context of quantum chemistry the coherent state \eqref{eqn:thouless-cs-1}
is designated as the \textit{Thouless parametrization} of a Slater
determinant. \cite{Thouless1960} Here, the label $z$ is understood as an
$M\times N$ matrix, the number of degrees of freedom of the corresponding phase
space being $d=M \times N$. The elements $z_{\mu\alpha}$ describe the mixing
between occupied and virtual spin orbitals in such a way that any
single-determinantal state not orthogonal to $\ket{\Phi_0} = \hat{A}_N
\prod^{N}_{\alpha=1} \ket{\phi^\bullet_\alpha}$ can be represented by
$\ketn{z}$. \cite{NoteAdded}

The overlap between two distinct Thouless-parame-trized Slater determinants is
readily found to be:
\begin{equation}\label{eqn:thouless-overlap}
	\braketn{z}{z'}
	=\det (I_N +z^\dagger z'),
\end{equation}
where $I_N$ is the identity matrix of size $N \times N$. Following the
definitions given in \S\ref{sec:cs-formalism}, geometrical properties are
determined from the function:
\begin{equation}\label{eqn:thouless-f}
	f(z^\conj, z') =
	\log [ \det (I_N +z^\dagger z') ].
\end{equation}
From \eqref{eqn:g-def}, one finds (through elementary determinant and matrix
identities) that the phase-space metric can be expressed as:
\begin{equation}\label{eqn:thouless-g}
	g_{\mu\nu,\alpha\beta}(z^\conj ,z)
	=[(I_M +z z^\dagger )^{-1}]_{\nu\mu}[(I_N+z^\dagger z)^{-1}]_{\alpha\beta},
\end{equation}
where $I_M$ is the $M \times M$ identity matrix and the derivatives of $f$ were
taken with respect to $z_{\mu\alpha}$ and $z^\conj_{\nu\beta}$. Despite the
complicated outlook of \eqref{eqn:thouless-g} the classical equations of
motion for the Thouless parameters $z$, as governed by a standard
many-body Hamiltonian consisting of one- and two-body terms, display a simple
structure -- see, for instance, Section IIIA in Ref.~\onlinecite{Deumens1994}.

Finally, the integration measure appearing in the closure relation
\eqref{eqn:closure} is found to be:
\begin{equation}\label{eqn:thouless-measure}
	\dif\mu(z^\conj, z)
	=\kappa \:[\det(I_N+ z^\dagger z)]^{-n}
	\prod^{N}_{\alpha=1} \prod^{M}_{\mu=1} \frac{\dif^2 z_{\mu\alpha}}{\pi}.
\end{equation}
The normalization constant can be computed by evaluating the required
phase-space integral through a recurrence relation method; the result is $\kappa
= \prod^{N}_{\gamma=1} \frac{(n +1 -\gamma)!}{(N +1 -\gamma)!}$.

As one might infer from the above, fermionic coherent states of this kind are
somewhat more intricate than the ones discussed in the previous examples.
And, although the content of the present paper encompasses all basic
ingredients needed to implement the proposed method in terms of these states, a
proper treatment would nevertheless require introduction of additional concepts.
Therefore, we do not pursue applications involving this particular set of
coherent states in this paper.

\section{Generalized CCS method}
\label{sec:ccs-generalized}

The coupled coherent states (CCS) method, as originally developed by
Shalashilin and Child \cite{Shalashilin2000, Shalashilin2001b, Shalashilin2004a}
using canonical coherent states, belongs to the family of multiconfigurational
guided-basis methods. Its characteristic attributes are the non-orthogonality of
the basis set and the use of simple classical mechanics \cite{Note07} to guide
the basis elements, as opposed to more complicated full-variational approaches
like the Gaussian-based version of the multiconfigurational time-dependent
Hartree (G-MCTDH) method. \cite{Burghardt1999, Worth2003a, Shalashilin2008}

Formulation of the method for Gaussian states is fairly straightforward and the
same is true in the generalized context. In order to better appreciate the
additional features that arise in the latter case, we will first present the
method in its continuum form; the discrete version is developed subsequently.

\subsection{The continuum version}
\label{sub:ccs-continuum}

We begin by considering the coherent-state decomposition of an arbitrary
quantum state,
\begin{equation}\label{eqn:ccs-ansatz-1}
	\ket{\psi}
	=\int\dif\mu(z) \ket{z} \braket{z}{\psi}
	=\int\dif\mu(z_0) \ket{z} \braket{z}{\psi},
\end{equation}
which follows from \eqref{eqn:closure}. It is assumed that $z = z(t)$ is bound
to obey the classical equations of motion \eqref{eqn:z-dot-1}. By virtue of
phase-space volume conservation, we are allowed to transfer the integration
measure to the initial instant and conveniently integrate over initial
conditions $z_0 = z(0)$, as indicated in the second equality in
\eqref{eqn:ccs-ansatz-1}. The derivation of the CCS equations amounts to finding
a solution of the \Schrodinger equation
\begin{equation}\label{eqn:schrodinger}
	\im\hbar \ket{\dot{\psi}}
	=\hat{H} \ket{\psi},
\end{equation}
for $\ket{\psi}$ in the form given by \eqref{eqn:ccs-ansatz-1} with the
\textit{ansatz}:
\begin{equation}\label{eqn:a-factor-1}
	\braket{z}{\psi}= C(z) e^{\ihbar S(z)},
\end{equation}
where $S(z)$ is the action defined in \eqref{eqn:action-0}. In other words, we
seek an equation of motion for the time-dependent amplitude $C(z)$ that solves
\eqref{eqn:schrodinger}. Let us make a few observations regarding this
particular choice of solution.

First, all quantities that specify $\ket{\psi}$ -- i.e.~$\ket{z}$, $C(z)$ and
$S(z)$ -- are to be regarded as functions of the initial conditions $z_0$. Thus,
the method is conceived as an initial value representation from its onset.

Second, we note that it follows from \eqref{eqn:a-factor-1} that $C(z)$ depends
on the initial state $\ket{\psi_0} = \ket{\psi(0)}$ through the relation
$C(z_0)= \braket{z_0}{\psi_0}$. In numerical applications, the integral in
\eqref{eqn:ccs-ansatz-1} has to be approximated somehow; the typical procedure
is to sample initial conditions $z_0$ in phase space with the overlap modulus
$|\braket{z_0}{\psi_0}|$ playing the role of a weight function. Once the $z_0$'s
have been properly sampled, the values of the corresponding $C(z_0)$'s are
uniquely defined.

Third, the motivation behind the factorization of $\braket{z}{\psi}$ into a
complex amplitude times an action exponential comes from a general result of
semiclassical theory, according to which the classical action provides a
first-order approximation to the phase of the quantum state. Since this phase
accounts for most of the wavefunction's oscillatory behavior, $C(z)$ is expected
to present a rather smooth time dependence, thus facilitating numerical
treatment.

We now proceed to look for a differential equation for $C(z)$. Taking the time
derivative of \eqref{eqn:a-factor-1} and making use of the \Schrodinger
equation, we find (after rearranging terms):
\begin{equation}\label{eqn:a-dot-continuous-1}
	\im\hbar\:\dot{C}(z)
	=\left[\im\hbar\braket{\dot{z}}{\psi}
	+\bra{z}\hat{H}\ket{\psi}
	+L(z)\braket{z}{\psi}\right]e^{-\ihbar S(z)}.
\end{equation}

Next, we factor out $\ket{\psi}$ by separating the scalar products on the
right-hand side of the equation with the help of the closure relation $\hat{1} =
\int\dif\mu(z')\ket{z'}\bra{z'}$, with $z' = z'(t)$, which leads to
\begin{equation}\label{eqn:a-dot-continuous-2}
	\im\hbar\:\dot{C}(z)
	=\int\dif\mu(z'_0)\braket{z}{z'}
	\Delta^2 H(z^\conj, z') C(z') e^{\ihbar[S(z')-S(z)]}.
\end{equation}
Here we have already shifted the integration measure to the initial instant
[$z'_0 = z'(0)$] and replaced the $\braket{z'}{\psi}$ that appeared under the
integral sign for $C(z') e^{\ihbar S(z')}$.

The coupling $\Delta^2 H(z^\conj, z')$ in \eqref{eqn:a-dot-continuous-2} is
given by
\begin{equation}\label{eqn:d2h-1}
	\Delta^2 H(z^\conj, z')
	= \im\hbar \frac{\braket{\dot{z}}{z'}}{\braket{z}{z'}}
	+H(z^\conj, z') +L(z),
\end{equation}
where the non-diagonal matrix element
\begin{equation}\label{eqn:non-diagonal-h}
	H(z^\conj, z')
	=\frac{\bran{z}\hat{H}\ketn{z'}}{\braketn{z}{z'}}
	=\frac{\bra{z}\hat{H}\ket{z'}}{\braket{z}{z'}},
\end{equation}
is an analytical function of $z^\conj$ and $z'$ that can be directly obtained by
analytical continuation of the classical Hamiltonian \eqref{eqn:hamiltonian}.

As a last step, we express the first term in \eqref{eqn:d2h-1} as a function of
readily computable quantities. Since $\bra{z} = e^{-\frac{1}{2}f(z^\conj, z)}
\bran{z}$ we observe that:
\begin{equation*}\label{eqn:braket-zz-0}
	\frac{\braket{\dot{z}}{z'}}{\braket{z}{z'}}
	= \frac{\braketn{\dot{z}}{z'}}{\braketn{z}{z'}}
	-\frac{1}{2} \od{}{t}{} f(z^\conj, z).
\end{equation*}
The total time derivative of $f(z^\conj, z)$ is simply
\begin{equation*}\label{eqn:dfdt}
	\od{}{t}{} f(z^\conj, z)
	=\sum^d_{\alpha=1} \left[
	\pd{f(z^\conj,z)}{z^\conj_\alpha}{} \dot{z}^\conj_\alpha
	+\pd{f(z^\conj,z)}{z_\alpha}{} \dot{z}_\alpha \right],
\end{equation*}
while the remaining term involving $\bran{\dot{z}}$ can be written as
\begin{equation*}\label{eqn:braket-zz}
	\frac{\braketn{\dot{z}}{z'}}{\braketn{z}{z'}}
	=\sum^d_{\alpha=1} \pd{f(z^\conj, z')}{z^\conj_\alpha}{} \dot{z}^\conj_\alpha,
\end{equation*}
owing to the the analyticity of $\bran{z}$ on $z^\conj$. Hence, collecting
together the above results and making the necessary substitutions in
\eqref{eqn:d2h-1}, we find that the coupling may be expressed as:
\begin{multline}\label{eqn:d2h-2}
	\Delta^2 H(z^\conj, z')
	=H(z^\conj, z') -H(z^\conj, z)
	\\
	+\im\hbar \sum^d_{\alpha=1} \left[
	\pd{f(z^\conj,z')}{z^\conj_\alpha}{}
	-\pd{f(z^\conj,z )}{z^\conj_\alpha}{}\right] \dot{z}^\conj_\alpha;
\end{multline}
which is an analytic function on $z'$.

By integrating the equation of motion \eqref{eqn:a-dot-continuous-2} the
amplitudes at time $t$ can be determined from their initial values. Once the
amplitudes are known, we can reconstruct the quantum state with
\eqref{eqn:ccs-ansatz-1}, reproduced below in terms of $C(z)$:
\begin{equation}\label{eqn:ccs-ansatz-2}
	\ket{\psi}
	=\int\dif\mu(z_0) \ket{z} C(z) e^{\ihbar S(z)}.
\end{equation}

The integro-differential equation \eqref{eqn:a-dot-continuous-2} -- with
$\Delta^2 H(z^\conj, z')$ given by \eqref{eqn:d2h-2} -- relates directly to the
canonical coherent states version of the CCS method and shares its attractive
characteristics, namely: (i) in the semiclassical regime, according to the
reasons mentioned earlier, the amplitude $C(z)$ is expected to have a smooth
time dependence; (ii) because of the coherent-state overlap $\braket{z}{z'}$,
the $z'$ integral is localized around $z$; \cite{Note08} and (iii) the coupling
between amplitudes of different basis elements is not only sparse but also
non-diagonal, since the integrand is identically zero when $z'= z$.

As a final remark, we should note that, if one performs a series expansion of
$\Delta^2 H(z^\conj, z')$ for small $|z'-z|$, one finds that this series begins
with a second-order term. In the generalized coherent state case however, unlike
the specific situation for canonical coherent states, this does not coincide
with the second- and higher-order terms in the Taylor series of $H(z^\conj,z')$.

\subsection{The discrete version}
\label{sub:ccs-discrete}

Here we re-derive the CCS equations using a discrete set of coherent states as
basis, as opposed to the continuous set employed in the previous section. We
shall find that the discrete formulas do not differ from their analogue
expressions in the canonical coherent state case. This is due to the fact that
all information concerning distinct coherent-state geometries is codified in a
number of key quantities, namely: the overlap, the phase-space metric and
the classical equations of motion. Each of these quantities participates in the
same way in the working equations, regardless of the particular type of coherent
state being used -- incidentally, a most desirable feature for programming
purposes, for it means that the core subroutines of the method are essentially
independent of geometry. Nevertheless, for the sake of consistency, we must
review the discrete formulas, for they are the ones actually used in practice.

\subsubsection{Unitary propagation}
\label{subsub:ccs-discrete-unitary}

The first step towards a discrete unitary formulation of the generalized CCS
method is the assumption that one is able to write down a closure relation by
employing a finite number of basis elements as below:
\begin{equation}\label{eqn:discrete-closure}
	\hat{1}
	=\sum^M_{j=1} \sum^M_{k=1} \ket{z_j} \Lambda_{jk} \bra{z_k},
\end{equation}
where $M$ is the size of the basis set. \cite{Note09} It suffices that this
closure relation represents the identity operator only on the dynamically
accessible part of the phase space and that it holds only during the time
interval upon which the propagation takes place.

In order to properly represent the identity, the matrix $\Lambda$ in
\eqref{eqn:discrete-closure} must satisfy the relation:
\begin{equation}\label{eqn:lambda}
	\delta_{jk}
	=\sum_{l} \Omega_{jl} \Lambda_{lk},
\end{equation}
where
\begin{equation}\label{eqn:overlap-matrix}
	\Omega_{jk} = \braket{z_j}{z_k},
\end{equation}
is the \textit{overlap matrix}. This guarantees that we have a well-defined
discrete coherent-state decomposition:
\begin{equation}\label{eqn:ccs-discrete-1}
	\ket{\psi}
	=\sum_{j,k} \ket{z_j} \Lambda_{jk} \braket{z_k}{\psi}.
\end{equation}
In other words, $\Omega$ must be sufficiently well-conditioned so that
expressions involving its inverse $\Lambda$ remain numerically stable during the
entire propagation. Therefore, the basis-set initial conditions must be sampled
in such a way as to ensure that this requirement is fulfilled. One such
procedure, that results in a well-conditioned overlap matrix (at initial time),
is described in appendix \ref{sec:sampling}.

Yet, nothing prevents that an initially well-conditioned overlap matrix becomes
singular at some later time -- a notable weakness of time-dependent methods
formulated with non-orthogonal basis sets.\cite{Kay1989} In the event that
$\Omega$ becomes singular, one should take appropriate measures before resuming
the propagation. In this regard, a particularly interesting protocol has been
proposed by Habershon.\cite{Habershon2012a} The `singularity problem', though,
did not occur in the simple applications considered in this paper.

Upon these considerations, we introduce a discrete set of $M$ amplitudes $C_j =
C(z_j)$ as well as their corresponding action phases $S_j = S(z_j)$, writing:
\begin{equation}\label{eqn:ansatz-discrete}
	\braket{z_j}{\psi}
	= C_j e^{\ihbar S_j}.
\end{equation}
Next, we proceed exactly as in \S\ref{sub:ccs-continuum}. The equation of motion
in the discrete unitary case is then readily found to be:
\begin{equation}\label{eqn:a-dot-discrete-1}
	\im\hbar\:\dot{C}_j
	=\sum_{k,l} \Omega_{jk}\: \Delta^2 H_{jk} \:\Lambda_{kl} \:
	C_l\:e^{\ihbar(S_l-S_j)},
\end{equation}
with coupling matrix given by:
\begin{multline}\label{eqn:d2h-discrete}
	\Delta^2 H_{jk}
	=H(z^\conj_j, z_k) -H(z^\conj_j, z_j)
	\\
	+\im\hbar \sum^d_{\alpha=1} \left[
	\pd{f(z^\conj_j,z_k)}{z^\conj_{j\alpha}}{}
	-\pd{f(z^\conj_j, z_j)}{z^\conj_{j\alpha}}{}\right] \dot{z}^\conj_{j\alpha}.
\end{multline}

In practice, matrix $\Lambda$ is never explicitly constructed; rather, one
introduces a set of auxiliary amplitudes $D =(D_1, D_2, \ldots, D_{M})$ which
are related to the coefficients $C =(C_1, C_2, \ldots, C_{M})$ according to
\begin{equation}\label{eqn:d-amplitudes}
	\sum_{k} \Omega_{lk} D_k e^{\ihbar (S_k -S_l)}
	= C_l.
\end{equation}
Thus, at every time step $D$ is obtained from $C$ by means of the above
intermediate equation -- an operation that requires solving a linear system of
size $M$. Then, the equation of motion \eqref{eqn:a-dot-discrete-1} can be
recast as:
\begin{equation}\label{eqn:a-dot-discrete-2}
	\im\hbar\:\dot{C}_j
	=\sum_{k} \left[ \Omega_{jk}
	\:\Delta^2 H_{jk} \:e^{\ihbar(S_k-S_j)} \right] D_k,
\end{equation}
while the quantum state is expressed in terms of amplitudes $D$ as:
\begin{equation}\label{eqn:psi-discrete}
	\ket{\psi}
	=\sum_{k} \ket{z_k} D_k e^{\ihbar S_k}.
\end{equation}
The propagation scheme comprised by Eqs.~\eqref{eqn:z-dot-1},
\eqref{eqn:lagrangian-1} and \eqref{eqn:action-0}, together with
Eqs.~\eqref{eqn:d-amplitudes}, \eqref{eqn:a-dot-discrete-2} and
\eqref{eqn:psi-discrete}, represents the standard form of the generalized
coherent-state method proposed here. It can be shown to preserve normalization
-- given by the sum $\sum_k C^\conj_k D_k$ -- as long as the overlap matrix
remains sufficiently well-conditioned. In addition, it preserves total energy
(for time-independent Hamiltonians) as long as the identity operator can be
resolved in terms of the basis-set elements, though this situation is hardly
achieved in multidimensional problems. It has been pointed out that energy
conservation is closely related to the accuracy of the CCS method.
\cite{Habershon2012a} Thus, by monitoring total energy, one can make an
`on-the-fly' diagnosis as regards to the quality of the CCS results; indeed we
observe in our simulations that deviations from the exact quantum solution are
accompanied by fluctuations in total energy.

\subsubsection{Non-unitary case}
\label{subsub:ccs-discrete-non-unitary}

It may be convenient -- particularly when the system under study has only one or
two degrees of freedom -- to attempt a more straightforward discretization of
the closure relation, as below:
\begin{equation}\label{eqn:discrete-non-unitary-closure}
	\hat{1}
	\approx \sum^M_{k=1} \ket{z_k} \lambda_k \bra{z_k},
\end{equation}
with $\lambda_k$ approximating the integration measure $\dif\mu(z_k)$ at each
phase-space point.

The equation of motion for $C$ in this case can be obtained at once from
\eqref{eqn:a-dot-discrete-1} by setting $\Lambda_{kl} = \lambda_k \delta_{kl}$:
\begin{equation}\label{eqn:a-dot-discrete-non-unitary}
	\im\hbar\:\dot{C}_j
	=\sum_{k} \lambda_k \left[
	\Omega_{jk} \:\Delta^2 H_{jk} \:e^{\ihbar(S_k-S_j)} \right] C_k.
\end{equation}
Similarly, the quantum state in this case is given by:
\begin{equation}\label{eqn:psi-discrete-non-unitary}
	\ket{\psi}
	= \sum_k \lambda_k \ket{z_k} C_k e^{\ihbar S_k}.
\end{equation}
This propagation scheme is computationally less demanding than the one discussed
in the previous section  -- if the basis-set size is kept the same --, since
there is no need to solve a linear system at each time step. On the other hand,
a larger basis set (usually constructed as a grid in phase space) may be
necessary to converge the results if \eqref{eqn:a-dot-discrete-non-unitary} is
employed. Moreover, as discussed by Shalashilin and Child,
\cite{Shalashilin2004a} the direct discretization of equation
\eqref{eqn:a-dot-continuous-2} does not preserve the unitarity of an exact
quantum time evolution, that is, the norm of the propagated quantum state is not
automatically conserved, meaning that results must be normalized on output.

\subsection{Classical propagation}
\label{sub:classical-propagation}

To end this section, we discuss the particular situation whereupon a single
coherent-state basis element is used to describe the system:
\begin{equation}\label{eqn:psi-classical}
	\ket{\psi}
	= \ket{z} e^{\ihbar S(z)}.
\end{equation}
As can be seen from the above equation, this scheme only applies if the quantum
state to be propagated is a coherent state, that is: $\ket{\psi_0} = \ket{z_0}$.
The form of the approximated $\ket{\psi}$, with an action phase, can be derived
from the working equations of the previous sections by setting the basis-set
size $M = 1$, in which case we find that $\dot{C} = 0$ and hence $C(t) = C(0)
= 1$.

We shall denominate the primitive method defined by \eqref{eqn:psi-classical},
together with \eqref{eqn:z-dot-1}, \eqref{eqn:lagrangian-1} and
\eqref{eqn:action-0}, as the \textit{classical propagation scheme}, in view of
the fact that only classical ingredients are present in its formulation. It
serves as a reference method, against which more sophisticated approaches, such
as those described earlier, may be confronted -- which is useful, for example,
in order to identify non-classical behavior (defined in this sense), as in
\S\ref{sub:su3-triple-well}.

It should be pointed out that if the Hamiltonian of the system is such that
application of the time-evolution operator maps one coherent state onto another
-- or, more formally, when the Hamiltonian is an element of the Lie algebra
associated with the set of coherent states under consideration --, then the
classical propagation scheme actually gives the exact solution. In other
situations it may provide a reasonable approximation for very short times.

\section{Application examples}
\label{sec:ccs-application}

\subsection{Condensate in a double-well potential}
\label{sub:su2-double-well}

As a first application we consider a simplified model for the dynamics
of an $N$-particle Bose-Einstein condensate trapped in a double-well potential,
where individual bosons interact through contact forces, that is, with an
interacting potential $V(\mathbf{x}, \mathbf{x}') \propto \delta(\mathbf{x}
-\mathbf{x}')$. This model has been discussed in detail in several works;
\cite{Milburn1997, Viscondi2009, Viscondi2009a} here we briefly sketch its main
ideas.

In the two-mode approximation, it is assumed that only the single-particle
ground state and first excited state of the double-well potential, as obtained
from first-order perturbation theory, have a significant occupation, so that the
dynamics of the system is restricted to these two levels. This should be a good
approximation if one seeks to describe the low temperature regime, wherein most
of the particles are expected to be occupying the ground state -- see
Ref.~\onlinecite{Sakmann2009} for a discussion on the validity of the two-mode
approximation to the double-well problem.

Since the number of bosons is preserved, the particle number operator $\hat{N}$
is a constant of the motion. By means of a well-known homomorphism between the
algebra $\mathrm{su}(2)$ and the bosonic creation and annihilation operators,
the bosonic dynamics can be described in terms of three independent angular
momentum operators -- Schwinger's pseudo-spin operators \cite{Schwinger} -- with
total angular momentum $J = N/2$.

In terms of these operators the Hamiltonian of the model
is: \cite{Note10}
\begin{equation}\label{eqn:2well-hamiltonian}
	\hat{H}
	=\Omega \hat{J}_{x} +\frac{2\chi}{N-1} \hat{J}^{2}_{z},
\end{equation}
where the tunneling rate $\Omega$ equals the energy difference between the two
occupied single-particle states and the self-collision parameter $\chi$ is
proportional to the interaction strength of bosons located in the same potential
well.

Applying definitions \eqref{eqn:hamiltonian} and \eqref{eqn:su2-cs-1} to
\eqref{eqn:2well-hamiltonian}, the classical Hamiltonian becomes \cite{Note11}
\begin{equation}\label{eqn:2well-hamiltonian-3}
	H(z^\conj, z)
	=\frac{N \Omega}{2} \frac{z +z^\conj}{1 +z^\conj z}
	+\frac{N \chi}{2} \frac{(1 -z^\conj z)^2}{(1 +z^\conj z)^2},
\end{equation}
in which we discarded, without any loss, a constant term. From identity
\eqref{eqn:z-dot-1}, the equation of motion for $z(t)$ is:
\begin{equation}\label{eqn:2well-zdot}
	\im \dot{z}
	= \frac{\Omega}{2} (1 -z^2)
	+2 \chi \frac{z (z^\conj z -1)}{1 +z^\conj z}.
\end{equation}
Notice that the $(N-1)^{-1}$ scaling of the self-collision parameter in
\eqref{eqn:2well-hamiltonian} was specifically chosen so that the equation of
motion \eqref{eqn:2well-zdot} is independent of particle number $N$, therefore
representing a well-defined classical limit of the system when $N \rightarrow
\infty$. This also means that, by employing the CCS method with a fixed set of
classical trajectories, we can obtain quantum solutions for different particle
number regimes.

\subsubsection{Non-unitary CCS with spin coherent states}
\label{subsub:su2-ccs-method}

We shall examine the quantum dynamics of the Hamiltonian
\eqref{eqn:2well-hamiltonian} from the perspective of a non-unitary
implementation of the CCS~method. Following the procedure outlined in
\S\ref{subsub:ccs-discrete-non-unitary}, a direct discretization of expressions
\eqref{eqn:a-dot-continuous-2} and \eqref{eqn:ccs-ansatz-2} is performed by
replacing the integrals over phase-space variables at $t = 0$ by finite
summations over a discrete set of initial conditions.

The accuracy of this straightforward discretization procedure is strongly
dependent on the selection of initial values over the phase space. In the
particular case of a regular grid, results are found to be extremely sensitive
to characteristics such as spacing, placement, number of points and, most
importantly, the choice of grid variables.

For the subsequent numerical calculations, we consider a regular grid in the
phase-space coordinates $(\eta,\zeta)$, which are related to the
$\mathrm{SU}(2)$ angular coordinates $(\theta,\phi)$ defined in equation
\eqref{eqn:su2-theta-phi}, according to:
\begin{align}\label{eqn:2well-new-coordinates}
	\eta
	&= \theta,
	\\
	\zeta
	&= \phi\sin\theta.
\end{align}
This choice aims at reducing the inherent difficulties imposed by the spherical
topology of spin coherent states at $\theta \approx 0\;\mathrm{or}\;\pi$.

In terms of this new set of coordinates, the integration measure
\eqref{eqn:su2-measure-theta-phi} takes a simpler form:
\begin{equation}\label{eqn:2well-measure-alpha-beta}
	\dif\mu
	=\frac{2J +1}{4 \pi} \dif \eta \dif \zeta,
\end{equation}
which, unlike expressions \eqref{eqn:su2-measure} and
\eqref{eqn:su2-measure-theta-phi}, is independent of the integration variables.
Hence, the $\lambda_k$'s that appear in the non-unitary evolution equation
\eqref{eqn:a-dot-discrete-non-unitary} and also in
\eqref{eqn:psi-discrete-non-unitary} are all equal to the same time-independent
$\lambda$ given by:
\begin{equation}\label{eqn:2well-lambda}
	\lambda
	=\frac{2J +1}{4 \pi} \Delta \eta \Delta \zeta,
\end{equation}
where $\Delta\eta$ and $\Delta\zeta$ are the grid spacings in the $\eta$ and
$\zeta$ directions, respectively.

\begin{figure}[htb]
\begin{center}
\includegraphics[width=0.48\textwidth]{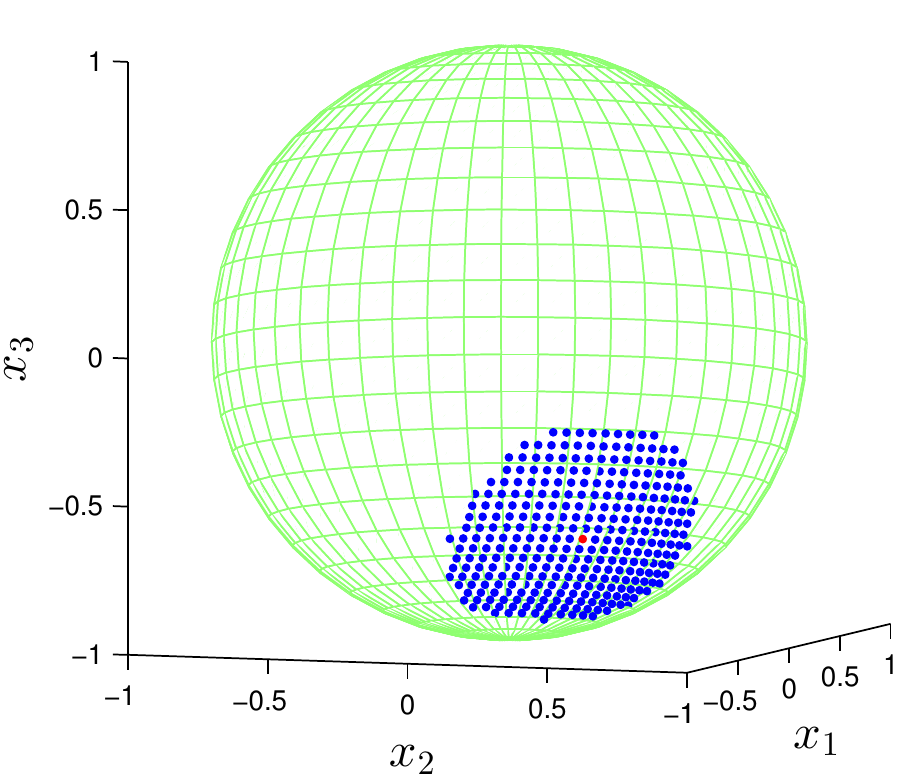}
\includegraphics[width=0.48\textwidth]{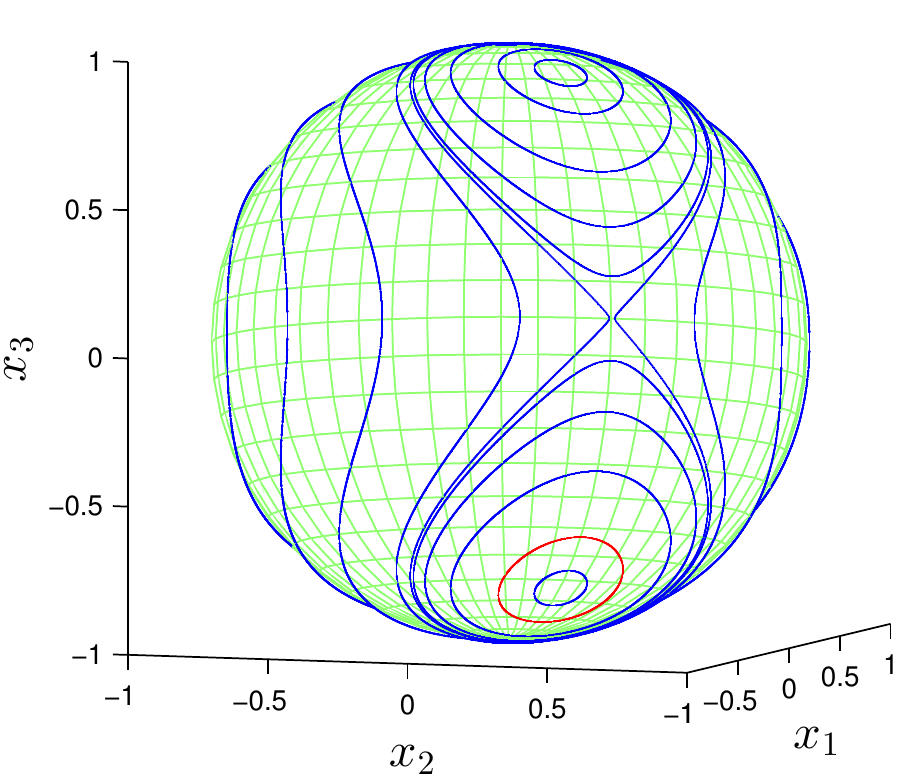}
\caption{\label{fig:initial-grid}
Upper panel: regular grid of classical initial conditions with approximately
three hundred initial values for the case of $N = 100$ particles. The red dot
indicates the center of the grid, whose angular coordinates are $(\theta',
\phi') =(\smallfrac{\pi}{4}, 0)$. This phase-space point represents the label of
the initial coherent state $\ket{\psi_0} = \ket{z'}$. Bottom panel: some
examples of classical trajectories over the spherical phase space. The red orbit
corresponds to the initial value $z'$. The Cartesian coordinates used to plot
these trajectories are related to the coherent-state angular variables according
to $(x_1, x_2, x_3) = (\sin\theta\cos\phi, \sin\theta\sin\phi, -\cos\theta)$.}
\end{center}
\end{figure}

The initial quantum state to be propagated is chosen to be the spin coherent
state $\ket{\psi_{0}} =\ket{z'}$ with
\begin{equation*}\label{eqn:2-well-initial-state}
	z'
	=\tan(\pi/8),
\end{equation*}
specified by angular coordinates $(\theta',\phi')=(\smallfrac{\pi}{4},0)$. As it
is usually assumed in most guided-basis methods, the most important dynamical
contributions -- at least for short-time propagation -- are expected to arise
from classical trajectories initially located at the same phase-space region
occupied by $\ket{\psi_0}$. Therefore, the grid of initial conditions is
intentionally centered at the point $(\theta',\phi')$, as illustrated on the
upper panel of Fig. \ref{fig:initial-grid} for the particular case of $N = 100$
particles -- the bottom panel portrays the classical dynamics in phase space.
Notice that the classical trajectory with initial value $z'$ is located close to
a stable equilibrium point and within the boundaries of a separatrix of motion.

With the purpose of quantifying the agreement between the CCS-propagated quantum
state $\ket{\psi_{\mathrm{ccs}}(t)}$ and the exact result
$\ket{\psi_{\mathrm{exact}}(t)}$ -- obtained via diagonalization of
\eqref{eqn:2well-hamiltonian} in the angular momentum basis -- we shall compute
the \textit{fidelity}:
\begin{equation}\label{eqn:fidelity-def}
	\mathcal{F}(t)
	=|\braket{\psi_{\mathrm{exact}}(t)}{\psi_{\mathrm{ccs}}(t)}|.
\end{equation}
The fidelity reflects the physical similarity between two states: its values are
restricted to the interval $0 \leq \mathcal{F} \leq 1$, with the maximum value
corresponding to physically identical quantum states.

We wish to analyze the behavior of $\mathcal{F}(t)$ as the number of particles
in the system is changed. There is a subtlety involved, though: due to its
fundamental property of minimal uncertainty, \cite{Delbourgo77a, Delbourgo77b}
the coherent state $\ket{z}$ represents a quantum state with maximal
localization in phase space around the point $z$. Moreover, for spin coherent
states, the linear dimensions of the phase-space region effectively occupied by
$\ket{z}$ decrease proportionally to $1/\sqrt{N}$. \cite{Note12} Thus, in order
to make a meaningful comparison amongst trajectory-based propagations carried
out at different particle number regimes, one must account for the `shrinking'
of the initial state $\ket{\psi_0} = \ket{z'}$ as $N$ grows larger. Therefore,
in our simulations, the grid spacing was reduced proportionally to $1/\sqrt{N}$
whilst the number of initial conditions in each run was roughly unchanged.

\begin{figure}[hbt]
\begin{center}
\includegraphics[width=0.48\textwidth]{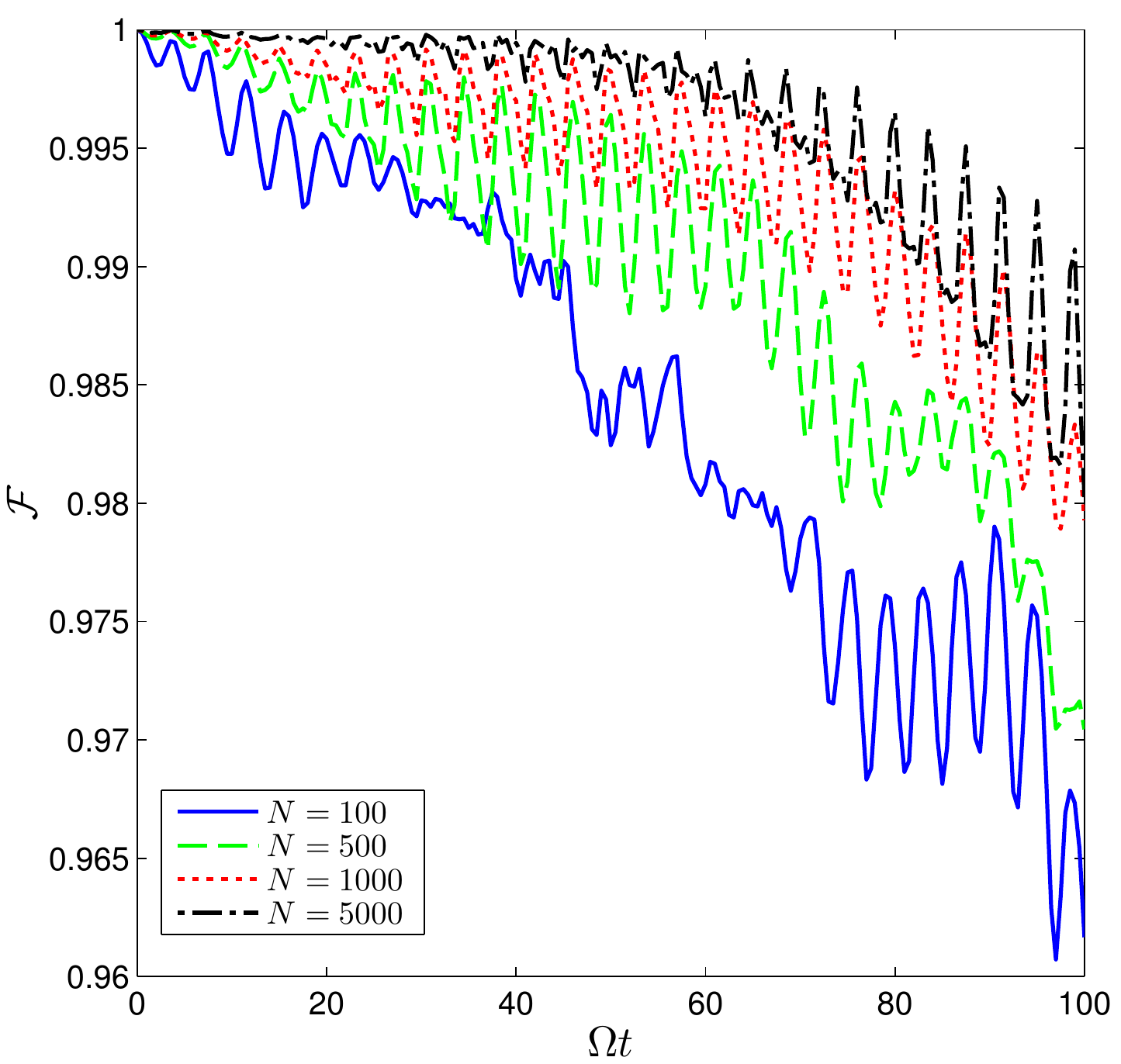}
\caption{\label{fig:fidelity}
Fidelity as a function of the dimensionless time $\Omega t$ with parameters
$\Omega =1.0$, $\chi =1.0$ for different particle number regimes. Each run
employed roughly three hundred trajectories -- the number of points varies
slightly for different runs due to the cropping of grid borders.}
\end{center}
\end{figure}

Fig.~\ref{fig:fidelity} shows $\mathcal{F}$ as a function of time for various
values of $N$. Notice that, even for long times, the non-unitary results remain
accurate ($\mathcal{F} >0.96$). On the other hand, it is clear that the fidelity
tends to decrease over time, evidencing the accumulation of numerical errors in
the CCS dynamics. In part, this inaccuracy stems from the non-unitary nature of
the discretization procedure. However, to a large extent, numerical error arises
due to dispersion of the classical trajectories over phase space, resulting in
an incomplete description of the quantum system.

Also note that the fidelity is consistently higher for larger particle numbers.
Since the number of classical trajectories was kept nearly constant for all the
runs, and the range of the initial condition grid was adjusted so as to reflect
the width of the initial quantum state, we may conclude that the CCS method
is better suited for describing the dynamics in the \textit{many-particle
regime}, which we recognize as the \textit{semiclassical regime} for this
problem. Indeed, it is well known that the limiting case $N \rightarrow \infty$
(or $J \rightarrow \infty$, as it can be interpreted in this problem) coincides
with the classical limit of quantum mechanics. \cite{Yaffe82}

In this sense, the CCS method, though in principle formulated as an exact
method, when numerically implemented -- and therefore subjected to inevitable
practical limitations -- should be regarded more as a semiclassical technique
than a genuine quantum approach, an assertion that can be made with respect to
both its non-unitary and unitary versions.

Finally, we point out that, for the system under consideration, the
computational cost of the method is insensitive to particle number. Hence, for
very large values of $N$, the computational resources needed for evaluating the
exact quantum dynamics by diagonalization of the Hamiltonian
\eqref{eqn:2well-hamiltonian} will certainly exceed the analogous requirements
of the non-unitary CCS method.

\subsection{Condensate in a triple-well potential}
\label{sub:su3-triple-well}

Next, we consider a simplified model describing an $N$-particle
Bose-Einstein condensate trapped in a symmetric triple-well potential where
individual bosons are once more assumed to interact by contact forces; the main
ideas involved are as follows: the triple-well trapping
potential, under suitable conditions, can be approximated by an harmonic
expansion around each of its three (symmetrically located) minima. The
three-fold degenerate fundamental states of this approximated problem can be
determined without difficulty. It is then assumed that the dynamical regime is
such that the energy eigenspace spanned by these three local modes
is sufficiently isolated from the rest of the single-particle spectrum, so
that at low temperatures they alone provide an adequate description of the
system. For more details on the derivation and particularities of this
model, see Refs. \onlinecite{Viscondi2010, Viscondi11a}.

Let $a_1$, $ a_2$ and $a_3$ denote the annihilation operators associated with
the aforementioned fundamental single-particle modes related to the
locally approximated wells. In terms of these bosonic operators, the
`three-mode approximation' \cite{Viscondi11a} to the Hamiltonian is:
\begin{equation}\label{eqn:3-well-hamiltonian-1}
	\hat{H}
	=\Omega \sum_{1 \leq \alpha\neq\beta \leq3} a^\dagger_\alpha a_\beta
	+\frac{\chi}{N -1} \sum_{1 \leq\gamma\leq 3}
	a^\dagger_\gamma a^\dagger_\gamma a_\gamma a_\gamma,
\end{equation}
where $\Omega$ is the tunneling rate, describing hops between adjacent wells,
and $\chi$ is the collision parameter, that controls the strength of two-body
interactions within the same well. \cite{Note13}

Owing to particle number conservation, this system is suitably described in
terms of $\mathrm{SU}(3)$ bosonic coherent states $\ket{z} = \ket{z_1,
z_2}$, which represent a particular case of the coherent states discussed
in \S\ref{subsub:sun-cs}. Using definition \eqref{eqn:sun-cs-1} together
with \eqref{eqn:3-well-hamiltonian-1}, we find from \eqref{eqn:hamiltonian} that
the classical Hamiltonian is
\begin{multline}\label{eqn:3-well-hamiltonian-2}
	H(z^\conj, z)
	=N \Omega \frac{(z^\conj_1 z_2 +z^\conj_2 z_1 +z^\conj_1
	+z_1 +z^\conj_2 + z_2)}{1 +z^\conj_1 z_1 + z^\conj_2 z_2}
	\\
	+N \chi \frac{(z^\conj_1 z_1)^2 +(z^\conj_2 z_2)^2 + 1}
	{(1 +z^\conj_1 z_1 +z^\conj_2 z_2)^2}.
\end{multline}
From \eqref{eqn:z-dot-1} it follows that the equations of motion are:
\begin{subequations}\label{eqn:3-well-zdot}
\begin{align}
	\im \dot{z}_1
	&= \Omega (1 +z_1 +z_2)(1 -z_1)
	-\frac{2 \chi z_1 (1 -|z_1|^2)}{1 +|z_1|^2 +|z_2|^2},
	\label{eqn:3-well-zdot-a}
	\\
	\im \dot{z}_2
	&= \Omega (1 +z_1 +z_2)(1 -z_2)
	-\frac{2 \chi z_2 (1 -|z_2|^2)}{1 +|z_1|^2 +|z_2|^2}.
	\label{eqn:3-well-zdot-b}
\end{align}
\end{subequations}
Similarly to the double-well model, we have deliberately tuned the
collision parameter $\chi$ with a $(N-1)^{-1}$ factor, thereby making the
classical dynamics independent of particle number; in this way the classical
system is well-defined in the limit $N\rightarrow \infty$.

\subsubsection{Unitary CCS with \texorpdfstring{$\mathrm{SU}(3)$}{SU(3)}
bosonic coherent states}
\label{subsub:su3-ccs-method}

The classical system defined in \eqref{eqn:3-well-zdot} has three dynamically
equivalent invariant subspaces, specified by the constraints: $z_1 = z_2$, $z_1
= 1$ and $z_2 = 1$. Let us concentrate on the first of these ($z_1 = z_2$) and
refer to it as the $\Gamma_1$ subspace.

Now, consider the set of operators $b_1$, $b_2$ and $b_3$, defined by the
canonical transformation:
\begin{subequations}\label{eqn:3-well-b}
\begin{align}
	b_1
	&= \smallfrac{1}{\sqrt{2}} (a_1 +a_2),
	\label{eqn:3-well-b1}
	\\
	b_2
	&= a_3,
	\label{eqn:3-well-b2}
	\\
	b_3
	&= \smallfrac{1}{\sqrt{2}} (a_1 -a_2).
	\label{eqn:3-well-b3}
\end{align}
\end{subequations}
It can be demonstrated that $\Gamma_1$ is an $\mathrm{SU}(2)$ subspace whose two
single-particle modes are associated with operators $b_1$ and $b_2$.
\cite{Viscondi11a} As a consequence, under the \textit{classical approximation}
described in \S\ref{sub:classical-propagation}, any $\mathrm{SU}(3)$ coherent
state initially located in $\Gamma_1$ will always display zero occupation of the
$b_3$ mode; as a matter of fact, the expectation value
\begin{equation}\label{eqn:3-well-bb-mean}
	\bra{z} b^\dagger_3 b_3 \ket{z}
	=\frac{N}{2}
	\frac{(z^\conj_1 -z^\conj_2) (z_1 -z_2)}
	{1 +z^\conj_1 z_1 +z^\conj_2 z_2},
\end{equation}
is identically null when $z_1 = z_2$.

This conclusion, however, does not apply to the actual quantum problem: even if
the initial state $\ket{\psi_0}$ has null occupation in the $b_{3}$ mode, this
situation will not be preserved as the system evolves in time. It is precisely
this `non-classical behavior' that we wish to describe using the
$\mathrm{SU}(3)$ CCS method.

Let us then consider the following initial state $\ket{\psi_0} = \ket{z'_1,
z'_2}$, with
\begin{equation*}\label{eqn:3-well-initial-state}
	z'_1 = z'_2 = \smallfrac{1}{\sqrt{2}} \tan(\pi/8),
\end{equation*}
and thus located on the classical invariant subspace $\Gamma_1$. We shall
propagate this state employing the unitary method developed in
\S\ref{subsub:ccs-discrete-unitary} and compute the occupation $Q(t)
=\bra{\psi(t)} b^\dagger_3 b_3 \ket{\psi(t)}$ of the single-particle mode
associated with $b_3$. Following \eqref{eqn:psi-discrete}, this function will be
calculated according to:
\begin{equation}\label{eqn:3-well-Q}
	Q
	=\sum_{j,k} D^\conj_j D_k \Omega_{jk}
	\left[ \frac{\bran{z_j} b^\dagger_3 b_3 \ketn{z_k}}{\braketn{z_j}{z_k}}
	\right] e^{\im (S_k -S_j)},
\end{equation}
where the non-diagonal matrix elements between square brackets can be obtained
by analytic continuation of \eqref{eqn:3-well-bb-mean}. We shall also monitor
the total energy of the system,
\begin{equation}\label{eqn:3-well-E}
	E
	=\sum_{j,k} D^\conj_j D_k \Omega_{jk} H(z^\conj_j, z_k) e^{\im (S_k -S_j)},
\end{equation}
as a means to probe the quality of our results.

In order to construct the basis set, it is necessary to choose adequate sampling
variables. In the present case we opt for angular variables $(\theta_1, \phi_1,
\theta_2, \phi_2)$ defined by:
\begin{align}\label{eqn:3-well-theta-phi}
	z_1
	&= \tan(\theta_1/2) e^{-\im\phi_1},
	\\
	z_2
	&= \tan(\theta_2/2) e^{-\im\phi_2}.
\end{align}
The initial conditions $z(0)$ are then randomly sampled around $z' = (z'_1,
z'_2)$ from Gaussian probability distributions expressed in terms of these
angular variables, with each angle being individually selected: for example,
$\theta_1(0)$ and $\phi_1(0)$ are selected according to
\begin{equation}\label{eqn:3-well-distribution-examples}
	\sim e^{-[\theta_1(0) -\theta'_1]^2 / 2 \sigma_{\theta_1}}
	;\ %
	\sim e^{-[\phi_1(0) -\phi'_1]^2 / 2 \sigma_{\phi_1}}.
\end{equation}
Notice that the widths of these distributions are adjustable parameters of the
method. The actual sampling procedure -- which also comprises specific criteria
for accepting and neglecting candidate basis elements -- is somewhat involved
and further details are reserved to appendix \ref{sec:sampling}. Once the
basis-set initial conditions are known, the amplitudes $C$ and $D$ of Eqs.
\eqref{eqn:ansatz-discrete} and \eqref{eqn:d-amplitudes} can be initialized
and propagation of \eqref{eqn:a-dot-discrete-2} may be started.

\begin{figure}[htb]
\begin{center}
\includegraphics[width=0.48\textwidth]{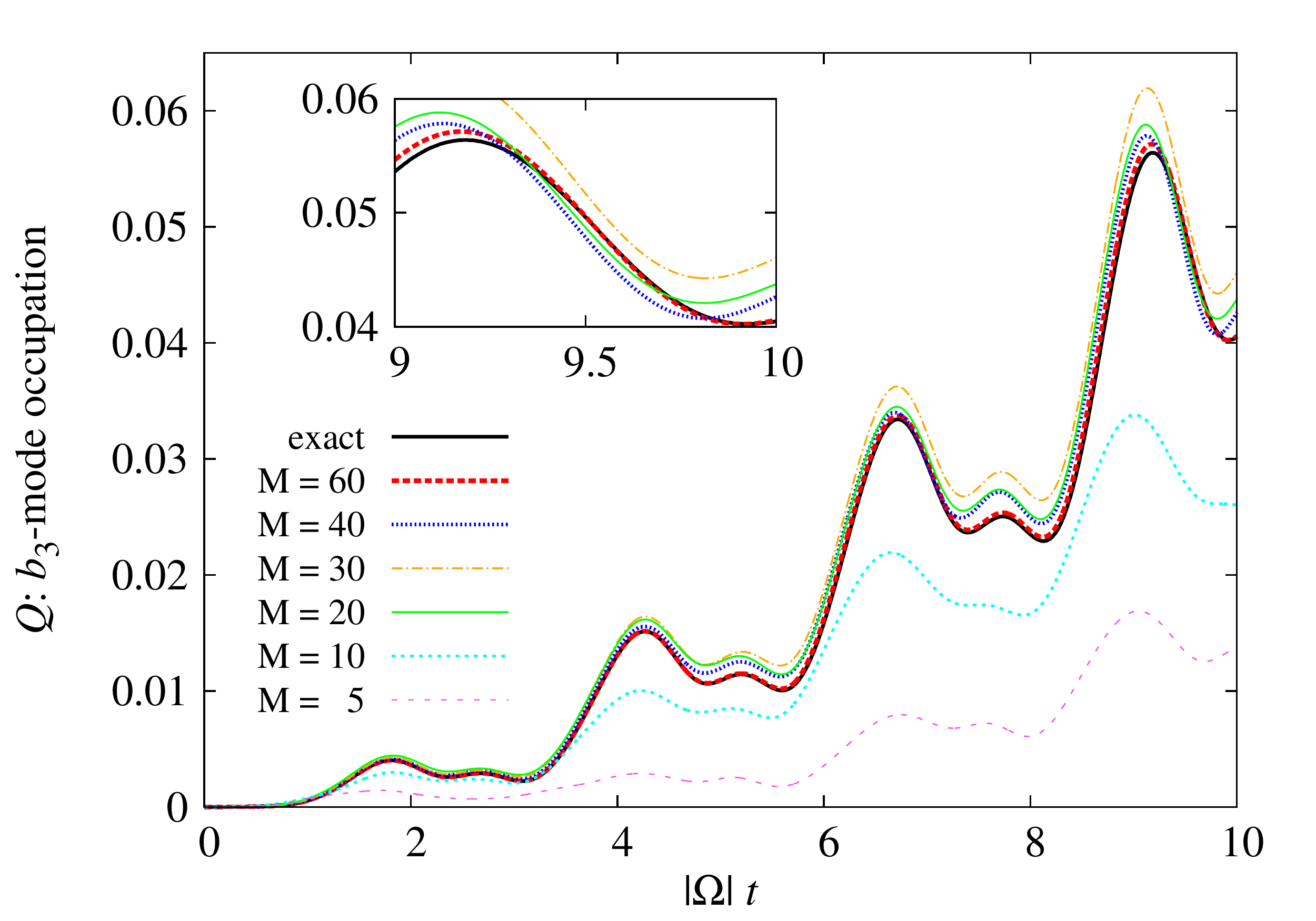}
\includegraphics[width=0.48\textwidth]{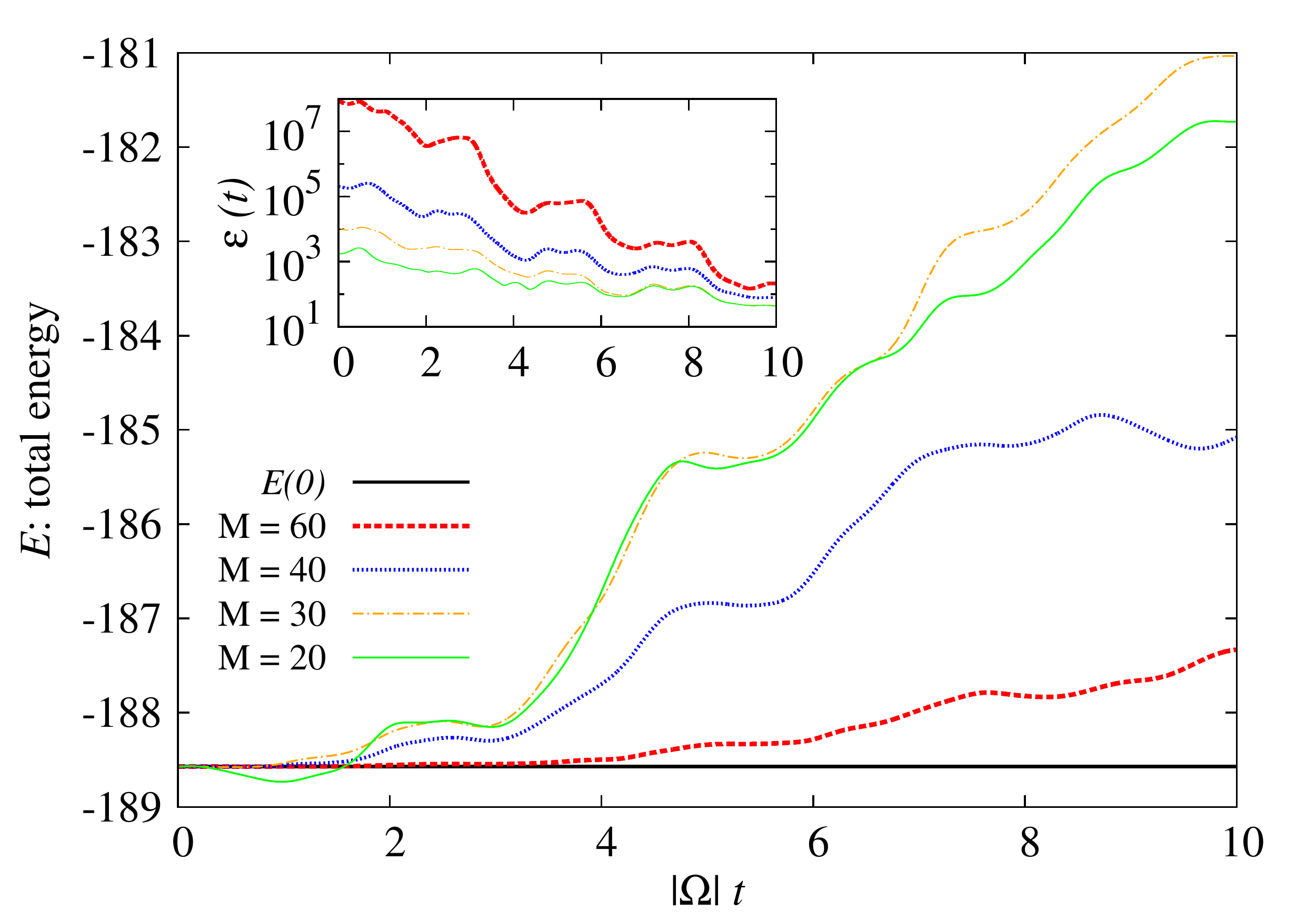}
\caption{Upper panel: Occupation of the classically inaccessible $b_{3}$ mode as
a function of dimensionless time $|\Omega|t$ for $N =100$, $\Omega = -1.0$,
and $\chi =-1.0$ and different basis-set sizes $M$; inset: zoom on the time
interval $9.0 \leq |\Omega|t \leq 10.0$. Bottom panel: Energy fluctuation as a
function of $|\Omega|t$ for the same runs shown on the upper panel; inset:
basis-set conditioning factor $\varepsilon(t)$ for the corresponding curves.}
\label{fig:b-occupation}
\end{center}
\end{figure}

We studied the case of $N =100$ trapped bosons, with tunneling rate $\Omega
=-1.0$, and collision parameter $\chi=-1.0$. The widths of the Gaussian
distributions -- exemplified in \eqref{eqn:3-well-distribution-examples} -- were
set to $\frac{\pi}{20}$ for both $\theta_1$ and $\theta_2$ and to
$\frac{\pi}{10}$ for both $\phi_1$ and $\phi_2$.

In order to visualize whether results converge as the basis-set size $M$ is
increased, simulations have been performed with different number of orbits while
maintaining all other parameters -- including sampling widths and random number
generator seed \cite{Note14} -- fixed.

Results obtained with the unitary $\mathrm{SU}(3)$ CCS method and by a direct
diagonalization of the Hamiltonian \eqref{eqn:3-well-hamiltonian-1} in the
bosonic Fock basis are compared in Fig.~\ref{fig:b-occupation}. On the upper
panel, the occupation of the classically inaccessible $b_3$ mode is displayed
for different basis-set sizes. On the bottom panel, the total energy expectation
value \eqref{eqn:3-well-E} for each run is shown as well as the corresponding
behavior of basis-set conditioning factors (see appendix~\ref{sec:sampling}).

It is found that CCS results improve as more basis elements are included in the
propagation, as expected -- although there is some anomalous behavior, with the
$M = 20$ run performing better than the $M = 30$ case. This is most likely due
to the basis-set sizes involved, which are probably too small for a uniform
convergence to be observed.

We also note that the quality of CCS results is intimately connected to
conservation of total energy: the more accurate runs are also those in which
energy is better conserved.

The most computationally expensive run, with $M = 60$ basis elements, led to an
initial basis-set conditioning factor $\varepsilon(0) \approx 8.8 \times 10^7$,
a value that diminished during the propagation. This behavior was observed in
all cases -- as shown in the inset of the bottom panel of Fig.
\ref{fig:b-occupation} --, indicating spread of the classical trajectories.
For this particular run ($M = 60$), excellent agreement between the CCS and
exact solutions is found, except for slight discrepancies at $|\Omega| t \gtrsim
8.0$.

It should be noted that the propagation with $M = 60$ -- which is quite a modest
number of basis elements for a problem with two degrees of freedom --, not only
proved to be accurate \cite{Note15} but also considerably faster than solving
the 5151-dimensional quantum eigensystem. This gain in efficiency, at least for
the model discussed here, drastically increases as more particles
are added to the system, since the dimension of the $\mathrm{SU}(3)$ bosonic
Fock space, given by $\frac{1}{2}(N+1)(N+2)$, grows rapidly with $N$. At the
same time, the system becomes more classical as more particles participate in
the dynamics, therefore making a trajectory-based approach more inviting.

Yet, despite these advantages, a reliable and positive way to check for the
convergence of the CCS method still remains to be developed -- the criterion of
energy non-conservation being only indicative of accuracy loss during
propagation. This represents a critical shortcoming when studying more complex
systems for which there are no exact numerical results -- or experimental data
-- available for comparison; for in those cases no precise statements about the
system could be made based on CCS results alone.

\section{Summary and conclusions}
\label{sec:conclusion}

In this work we have formulated a multiconfigurational, trajectory-guided
quantum propagation method whose distinctive quality consists in employing
generalized coherent states as basis elements. In this sense, the technique is
seen as a natural extension of the coupled coherent states method of
Shalashilin and Child \cite{Shalashilin2000, Shalashilin2001b, Shalashilin2004a}
whereupon frozen Gaussians are replaced by more general configurations; at the
same time, the main features of the original CCS are retained: quantum
amplitudes obey an integro-differential equation with sparse coupling and
present a smooth time dependence, owing to their oscillatory behavior being
partially compensated by the classical motion of the basis elements and their
action phases.

As pointed out in \S\ref{sec:cs-formalism}, no deep understanding of
group-theory concepts is necessary, neither to derive the basic equations of the
method nor to implement it numerically -- in fact, we have seen that all
geometrical quantities that enter the formulas can be straightforwardly
evaluated from the coherent-state overlap function alone and that the discrete
versions of the working equations do not differ in overall structure from their
analogue expressions in the original CCS approach.

In \S\ref{sec:ccs-generalized}, three versions of the method have been devised:
continuum, non-unitary and unitary. The continuum version most evidently
displays the novel elements due to the non-Euclidean geometry associated with
the generalized coherent-states and it serves primarily as a starting point for
a number of possible analytical approximations. The non-unitary version, in
turn, might be understood as a direct attempt to reproduce the continuum
formulas by reducing phase-space integrals into finite sums. The unitary
version, meanwhile, is the standard form of the method, being the most adequate
for the majority of practical applications. We have also briefly discussed one
weakness of this last version of the method which is the ill-conditioning of
the overlap matrix. In that respect, we note that the basis-set sampling
procedure that we propose -- detailed in appendix \ref{sec:sampling} -- is more
appealing than the techniques used in previous CCS applications, for it assures
an initially well-conditioned overlap matrix and hence stability of the
amplitude equation \eqref{eqn:a-dot-discrete-2}, at least for short time
propagation.

In \S\ref{sec:ccs-application}, we have illustrated the general aspects of the
proposed approach with applications to simple models of many-boson systems,
described in  terms of $\mathrm{SU}(2)$ and $\mathrm{SU}(3)$ bosonic
coherent states. One particular aspect, namely, the choice of appropriate
phase-space variables for either sampling of initial conditions or grid
construction, is found to be of great importance -- coordinates chosen for these
purposes must somehow reflect the intrinsic geometry of the coherent state under
consideration, otherwise only a poor representation of the system is obtained.
Moreover, the accuracy and efficiency of the method was established by comparing
results against exact quantum calculations. Excellent agreement was observed
and, in the triple-well system examined in \S\ref{sub:su3-triple-well}, small
discrepancies in the results were found to be correlated with energy
non-conservation. These tests also allowed us to gain some insight regarding the
domain of applicability of our formulas. A careful analysis of the fidelity in
the double-well system studied in \S\ref{sub:su2-double-well} revealed that the
CCS approach is best suited for describing the regime of large particle number,
also identified as the semiclassical regime of that problem. This conclusion can
in fact be extended for $\mathrm{SU}(n)$ bosonic systems in general. The same
goes for the observations concerning the computational advantages of the method
(over standard matrix diagonalization) when the number of particles is large;
assertions which were made within the context of the triple-well problem.

Though we have exemplified the use of the method with $\mathrm{SU}(2)$ and
$\mathrm{SU}(3)$ bosonic Hamiltonians, we emphasize that the formulation
presented is by no means limited to these types of systems -- in
addition to the $\mathrm{SU(n)}$ fermionic coherent states discussed in
\S\ref{subsub:thouless-cs}, other possible coherent-state parametrizations in
terms of which the method could be readily implemented were referenced
throughout \S\ref{sec:introduction}. Furthermore, the conclusion put forth in
\S\ref{sec:ccs-application}, namely, that propagation of quantum states with the
help of classically guided basis sets is better suited for describing systems in
their semiclassical regime, is certainly expected to hold for any coherent-state
representation that might be employed -- provided one correctly interprets what
`semiclassical regime' means in each case.

A few words regarding the purely computational aspects of the method are in
place. In the numerical calculations reported in this paper we have adopted the
simplest possible strategy of implementation, which consists in propagating
simultaneously the coherent-state trajectories and their corresponding
amplitudes. But since trajectories evolve independently, an alternative approach
would be to propagate them separately, storing the coherent-state coordinates at
pre-determined time intervals, and use this information afterwards in order to
propagate the fully coupled amplitude equation -- perhaps employing
interpolation algorithms if coordinates are needed at intermediate instants.

This preliminary propagation of classical orbits could, of course, take full
advantage of parallelization, especially if single basis elements are propagated
with ease. It is however in those cases in which the very propagation of
individual trajectories is a computationally demanding task -- either because
the system has an extremely large number of degrees of freedom or because no
analytical expression for the Hamiltonian is available -- that this `two-stage'
procedure would most definitely be the strategy of choice. In addition, it would
allow for more sophisticated sampling techniques, since it would then be
possible to select trajectories based on knowledge of their entire story, and
not relying just on their initial proximity to the quantum state. Thus, for
instance, one could identify orbits which, though unimportant at early stages of
the propagation, give a significant contribution later on.

As a final and interesting remark, we note that the multiconfigurational
Ehrenfest method, \cite{Shalashilin2010, Saita2012} as specifically designed for
`on-the-fly' non-adiabatic dynamics, can be obtained at once from the presented
formalism as the particular case wherein each basis-set element is taken to be a
composite coherent state consisting of a canonical part and an $\mathrm{SU}(n)$
part with the particle-number parameter $N$ set to unity. \cite{Note16} In this
picture, different `diabatic' potential surfaces would be represented by the $n$
single-particle states composing the $\mathrm{SU}(n)$ coherent state; further
extension to Born-Oppenheimer `adiabatic' energy surfaces could be achieved
without difficulty.

\section*{Acknowledgements}
This work was supported by FAPESP under project grants 2008/09491-9,
2011/20065-4, 2012/20452-0, and 2014/04036-2. In addition, MAMA acknowledges
support from CNPq. Numerical calculations made use of routines from the GNU
Scientific Library. \cite{GSLmanual}

\appendix
\section{Basis set sampling}
\label{sec:sampling}

The basis set sampling procedure that we propose for the unitary CCS method
(\S\ref{subsub:ccs-discrete-unitary}) is outlined here. It applies to any type
of coherent state once two geometry-dependent ingredients are provided: adequate
sampling coordinates $q = f(z)$ -- with a known inverse $z = f^{-1}(q)$ -- and a
weight distribution function $w(q)$ according to which these coordinates are to
be randomly selected. The procedure assumes that the initial state is a coherent
state, i.e.~$\ket{\psi_0} = \ket{z'}$, in which case the initial coherent-state
sampling coordinates $q' = f(z')$ must also be supplied.

The sampling strategy follows a very simple `one-by-one' protocol, which draws
inspiration from previously developed basis set conditioning techniques.
\cite{Habershon2012a} The procedure amounts to four steps:

(1) Take $\ket{z'}$ (the initial state itself) as the first basis element;

(2) Using the appropriate sampling coordinates $q$ and weight function $w(q)$,
randomly select a new basis element $z_j = f^{-1}(q_j)$ and temporarily add
$\ket{z_j}$ to the basis set;

(3) Compute the overlap matrix $\Omega$ and evaluate its conditioning factor
$\varepsilon = \lambda_{\mathrm{max}}/\lambda_{\mathrm{min}}$, where
$\lambda_{\mathrm{max}}$ and $\lambda_{\mathrm{min}}$ are the largest and
smallest eigenvalues of $\Omega$, respectively. \cite{Note17}

(4) If $\varepsilon$ is less than some threshold value
$\varepsilon_{\mathrm{lim}}$, accept $\ket{z_j}$ permanently adding it to the
basis set, whose size increases by 1. Else, discard the selected basis element,
in which case the basis-set size does not change. In either case, return to step
(2);

The above procedure is then iterated until either a desired basis-set size is
achieved or \textit{saturation} occurs, meaning that the algorithm is unable to
select a new $\ket{z_j}$ that satisfy the $\varepsilon$ threshold condition (a
certain maximum number of attempts may be specified). How fast
saturation takes place will depend upon the system's dimensionality, the
threshold value $\varepsilon_{\mathrm{lim}}$, the coherent-state parameters and
the details of the sampling distribution $w(q)$. Typically, we take
$\varepsilon_{\mathrm{lim}} \sim 10^{8}-10^{12}$, and use a predetermined
basis-set size below saturation, thus ensuring a reasonably well-conditioned
overlap matrix at initial time and hence the stability of the propagation (at
least for sufficiently short times).

This sampling protocol requires the eigenvalues of the overlap matrix to
be computed at every iteration. We point out, however, that this does not
compromise the method's efficiency since the sampling is performed only once,
before the actual propagation. Moreover, the overlap matrix
typically does not grow too large; this assertion holds even for
multidimensional systems, as long as the sampling distribution is kept
sufficiently localized around the initial-state coordinate $z'$ from where the
most relevant contributions to the initial value representation formula are
expected to originate. Finally, note also that the initial state $\ket{z'}$ is
always included in the basis set; this is crucial for accuracy of short-time
results and also secures that the initial norm is unity, regardless of how the
remaining basis elements are distributed in phase space.



\begin{thebibliography}{10}\label{sec:refs}

\bibitem{Makri1999}
N.~Makri.
\newblock {Time-dependent quantum methods for large systems.}
\newblock {\em Annual Review of Physical Chemistry}, 50(1):167, 1999.

\url {http://dx.doi.org/10.1146/annurev.physchem.50.1.167}

\bibitem{Miller1986}
W.~H. Miller and K.~A. White.
\newblock {Classical models for electronic degrees of freedom: The
  second-quantized many-electron Hamiltonian}.
\newblock {\em The Journal of Chemical Physics}, 84(9):5059, 1986.

\url {http://dx.doi.org/10.1063/1.450655}

\bibitem{Kirrander2011}
A.~Kirrander and D.~V. Shalashilin.
\newblock {Quantum dynamics with fermion coupled coherent states: Theory and
  application to electron dynamics in laser fields}.
\newblock {\em Physical Review A}, 84(3):033406, 2011.

\url {http://dx.doi.org/10.1103/PhysRevA.84.033406}

\bibitem{Note01}
Recently, Grossmann \textit{et.~al.} [F.~Grossmann, M.~Buchholz, E.~Pollak,
and M.~Nest. \newblock {Spin effects and the Pauli principle in semiclassical
electron dynamics}. \newblock {\em Physical Review A}, 89(3):032104, 2014. (\url
{http://dx.doi.org/10.1103/PhysRevA.89.032104})] have investigated, in a
semiclassical context, whether propagation with antisymmetrized basis states is
essential for the description of electron scattering.

\bibitem{Feldmeier2000}
H.~Feldmeier and J.~Schnack.
\newblock {Molecular dynamics for fermions}.
\newblock {\em Reviews of Modern Physics}, 72(3):655, 2000.

\url {http://dx.doi.org/10.1103/RevModPhys.72.655}

\bibitem{Kramer}
P.~Kramer and M.~Saraceno.
\newblock {\em Geometry of the Time-Dependent Variational Principle in Quantum
  Mechanics}.
\newblock Springer-Verlag, New York, 1981.

\url {https://dx.doi.org/10.1007/3-540-10579-4}

\bibitem{Delbourgo77a}
R.~Delbourgo.
\newblock Minimal uncertainty states for the rotation and allied groups.
\newblock {\em Journal of Physics A: Mathematical and General}, 10(11):1837,
  1977.

\url {http://stacks.iop.org/0305-4470/10/i=11/a=012}

\bibitem{Delbourgo77b}
R.~Delbourgo and J.~R. Fox.
\newblock Maximum weight vectors possess minimal uncertainty.
\newblock {\em Journal of Physics A: Mathematical and General}, 10(12):L233,
  1977.

\url {http://stacks.iop.org/0305-4470/10/i=12/a=004}

\bibitem{Zhang1990a}
W.-M. Zhang, D.~H. Feng, and R.~Gilmore.
\newblock {Coherent states: Theory and some applications}.
\newblock {\em Reviews of Modern Physics}, 62(4):867, 1990.

\url {http://dx.doi.org/10.1103/RevModPhys.62.867}

\bibitem{VanVoorhis2004}
T.~Van~Voorhis and D.~R. Reichman.
\newblock {Semiclassical representations of electronic structure and dynamics.}
\newblock {\em The Journal of Chemical Physics}, 120(2):579, 2004.

\url {http://dx.doi.org/10.1063/1.1630963}

\bibitem{Viscondi2015}
T.~F. Viscondi, A.~Grigolo, and M.~A.~M. de~Aguiar.
\newblock Semiclassical propagator in the generalized coherent-state
  representation.
\newblock Submitted, arXiv:1510.05952/quant-ph.

\url {http://arxiv.org/abs/1510.05952}

\bibitem{Note02}
Their discussion is based on a rough extension of Solari's semiclassical
propagators. See:

(a) H.~G. Solari. Glauber's coherent states and the semiclassical propagator.
\newblock {\em Journal of Mathematical Physics}, 27(5):1351, 1986.

\url {http://dx.doi.org/10.1063/1.527142}

(b) H.~G. Solari. Semiclassical treatment of spin system by means of coherent
states. \newblock {\em Journal of Mathematical Physics}, 28(5):1097, 1987.

\url {http://dx.doi.org/10.1063/1.527554}

A rigorous derivation of the generalized coherent-state propagator can be found
in a recent work. \cite{Viscondi2015}

\bibitem{Thouless1960}
D.~J. Thouless.
\newblock {Stability conditions and nuclear rotations in the Hartree-Fock
  theory}.
\newblock {\em Nuclear Physics}, 21:225, 1960.

\url {http://dx.doi.org/10.1016/0029-5582(60)90048-1}

\bibitem{Deumens1989a}
E.~Deumens and Y.~\"{O}hrn.
\newblock {Time-dependent dynamics of a determinantal state}.
\newblock {\em Journal of Molecular Structure (Theochem)}, 199:23, 1989.

\url {http://dx.doi.org/10.1016/0166-1280(89)80039-9}

\bibitem{Deumens1992}
E.~Deumens, A.~Diz, H.~Taylor, and Y.~\"{O}hrn.
\newblock {Time-dependent dynamics of electrons and nuclei}.
\newblock {\em The Journal of Chemical Physics}, 96(9):6820, 1992.

\url {http://dx.doi.org/10.1063/1.462571}

\bibitem{Deumens1994}
E.~Deumens, A.~Diz, R.~Longo, and Y.~\"{O}hrn.
\newblock {Time-dependent theoretical treatments of the dynamics of electrons
  and nuclei in molecular systems}.
\newblock {\em Reviews of Modern Physics}, 66(3):917, 1994.

\url {http://dx.doi.org/10.1103/RevModPhys.66.917}

\bibitem{Suzuki1983}
T.~Suzuki.
\newblock {Classical and quantum mechanical aspects of time-dependent
  Hartree-Fock trajectories}.
\newblock {\em Nuclear Physics A}, 398:557, 1983.

\url {http://dx.doi.org/10.1016/0375-9474(83)90302-0}

\bibitem{Kuratsuji1980a}
H.~Kuratsuji and T.~Suzuki.
\newblock {Path integral approach to many-nucleon systems and time-dependent
  Hartree-Fock}.
\newblock {\em Physics Letters B}, 92:19, 1980.

\url {http://dx.doi.org/10.1016/0370-2693(80)90293-2}

\bibitem{Kuratsuji1983}
H.~Kuratsuji and T.~Suzuki.
\newblock {Path integral approach to many-body systems and classical
  quantization of time-dependent mean field}.
\newblock {\em Progress of Theoretical Physics Supplements}, 74-75:209,
  1983.

\url {http://dx.doi.org/10.1143/PTPS.74.209}

\bibitem{Viscondi2011}
T.~F. Viscondi and M.~A.~M. de~Aguiar.
\newblock {Semiclassical propagator for SU(n) coherent states}.
\newblock {\em Journal of Mathematical Physics}, 52(5):052104, 2011.

\url {http://dx.doi.org/10.1063/1.3583996}

\bibitem{Viscondi2011b}
T.~F. Viscondi and M.~A.~M. de~Aguiar.
\newblock {Initial value representation for the SU(n) semiclassical
  propagator.}
\newblock {\em The Journal of Chemical Physics}, 134(23):234105, 2011.

\url {http://dx.doi.org/10.1063/1.3601344}

\bibitem{VanVoorhis2002}
T.~Van~Voorhis and E.~J. Heller.
\newblock {Nearly real trajectories in complex semiclassical dynamics}.
\newblock {\em Physical Review A}, 66(5):050501, 2002.

\url {http://dx.doi.org/10.1103/PhysRevA.66.050501}

\bibitem{VanVoorhis2003}
T.~Van~Voorhis and E.~J. Heller.
\newblock {Similarity transformed semiclassical dynamics}.
\newblock {\em The Journal of Chemical Physics}, 119(23):12153, 2003.

\url {http://dx.doi.org/10.1063/1.1626621}

\bibitem{Note03}
We note that the approximations to the generalized coherent-state path integral
considered by Kuratsuji and Suzuki \cite{Kuratsuji1983} -- as well as specific
formulations for Slater determinants \cite{Kuratsuji1980a, Suzuki1983}
-- are very much akin to the techniques develop in this paper.

\bibitem{Shalashilin2000}
D.~V. Shalashilin and M.~S. Child.
\newblock {Time dependent quantum propagation in phase space}.
\newblock {\em The Journal of Chemical Physics}, 113(22):10028, 2000.

\url {http://dx.doi.org/10.1063/1.1322075}

\bibitem{Shalashilin2001b}
D.~V. Shalashilin and M.~S. Child.
\newblock {Multidimensional quantum propagation with the help of coupled
  coherent states}.
\newblock {\em The Journal of Chemical Physics}, 115(12):5367, 2001.

\url {http://dx.doi.org/10.1063/1.1394939}

\bibitem{Shalashilin2004a}
D.~V. Shalashilin and M.~S. Child.
\newblock {The phase space CCS approach to quantum and semiclassical molecular
  dynamics for high-dimensional systems}.
\newblock {\em Chemical Physics}, 304:103, 2004.

\url {http://dx.doi.org/10.1016/j.chemphys.2004.06.013}

\bibitem{Note04}
The discussion is restricted to coherent states given in terms of analytical
complex parametrizations.

\bibitem{Note05}
Entries of the vector $z$ will be referenced by Greek letters $\alpha, \beta,
\gamma$.

\bibitem{Viscondi2013}
T.~F. Viscondi.
\newblock PhD thesis, Universidade Estadual de Campinas, Instituto de
  F\'{i}sica Gleb Wataghin, Campinas, S\~{a}o Paulo, Brazil, 2013.
\newblock Available in Portuguese
\href{http://www.bibliotecadigital.unicamp.br/document/?code=000904028&opt=4&lg}
{here.}

\bibitem{Note06}
In the majority of applications involving coherent states and classical
trajectories, the imaginary surface terms in $A$ are canceled out by
normalization factors. However, when \textit{duplicated phase-space}
trajectories are considered -- i.e.~when $z$ and $z^\conj$ are analytically
continued in such a way that they become \textit{completely independent}
variables (thereby doubling the number of degrees of freedom in the system) --,
the complex action $A$ plays a prominent role, for its analytical properties are
of the uttermost importance in those cases. \cite{Baranger01, Aguiar2010b,
Viscondi2011, Viscondi2011b, Viscondi2015}

\bibitem{Heller1981b}
E.~J. Heller.
\newblock {Frozen Gaussians: A very simple semiclassical approximation}.
\newblock {\em The Journal of Chemical Physics}, 75(6):2923, 1981.

\url {http://dx.doi.org/10.1063/1.442382}

\bibitem{Heller1991}
E.~J. Heller.
\newblock {Cellular dynamics: A new semiclassical approach to time-dependent
  quantum mechanics}.
\newblock {\em The Journal of Chemical Physics}, 94(4):2723, 1991.

\url {http://dx.doi.org/10.1063/1.459848}

\bibitem{Herman1984}
M.~F. Herman and E.~Kluk.
\newblock {A semiclassical justification for the use of non-spreading
  wavepackets in dynamics calculations}.
\newblock {\em Chemical physics}, 91:27, 1984.

\url {http://dx.doi.org/10.1016/0301-0104(84)80039-7}

\bibitem{Kluk1986}
E.~Kluk, M.~F. Herman, and H.~L. Davis.
\newblock {Comparison of the propagation of semiclassical frozen Gaussian wave
  functions with quantum propagation for a highly excited anharmonic
  oscillator}.
\newblock {\em The Journal of Chemical Physics}, 84(1):326, 1986.

\url {http://dx.doi.org/10.1063/1.450142}

\bibitem{Herman1994}
M.~F. Herman.
\newblock {Dynamics by semiclassical methods}.
\newblock {\em Annual Review of Physical Chemistry}, 45:83, 1994.

\url {http://dx.doi.org/10.1146/annurev.physchem.45.1.83}

\bibitem{Kay1994}
K.~G. Kay.
\newblock {Integral expressions for the semiclassical time-dependent
  propagator}.
\newblock {\em The Journal of Chemical Physics}, 100(6):4377, 1994.

\url {http://dx.doi.org/10.1063/1.466320}

\bibitem{Baranger01}
M.~Baranger, M.~A.~M. de~Aguiar, F.~Keck, H.~J. Korsch, and B.~Schellhaa\ss.
\newblock Semiclassical approximations in phase space with coherent states.
\newblock {\em Journal of Physics A: Mathematical and General}, 34:7227, 2001.

\url {http://stacks.iop.org/0305-4470/34/i=36/a=309}

\bibitem{Kay2006}
K.~G. Kay.
\newblock {The Herman-Kluk approximation: Derivation and semiclassical
  corrections}.
\newblock {\em Chemical Physics}, 322:3, 2006.

\url {http://dx.doi.org/10.1016/j.chemphys.2005.06.019}

\bibitem{Aguiar2010b}
M.~A.~M. de~Aguiar, S.~A. Vitiello, and A.~Grigolo.
\newblock {An initial value representation for the coherent state propagator
  with complex trajectories}.
\newblock {\em Chemical Physics}, 370:42, 2010.

\url {http://dx.doi.org/10.1016/j.chemphys.2010.01.020}

\bibitem{Arecchi72}
F.~T. Arecchi, E.~Courtens, R.~Gilmore, and H.~Thomas.
\newblock Atomic coherent states in quantum optics.
\newblock {\em Physical Review A}, 6(6):2211, 1972.

\url {http://dx.doi.org/10.1103/PhysRevA.6.2211}

\bibitem{Note07}
Owing to the finite width of the Gaussian functions employed as basis, CCS
trajectories actually evolve under a smoothed-out classical Hamiltonian whose
width-dependent terms are usually referred to as `simple quantum corrections'.

\bibitem{NoteAdded}
SU(n) fermionic coherent states should be contrasted with the
antisymmetrized Gaussian wavepackets mentioned in \S\ref{sec:introduction}.
While the former are defined in terms of a finite set of underlying molecular
spin-orbitals, thus providing a convenient description of the electronic
structures of chemical systems, the latter are more adequate for representing
otherwise localized fermions, such as protons and neutrons in the atomic
nucleus, \cite{Feldmeier2000} or perhaps outer shell electrons in highly excited
atoms. \cite{Kirrander2011}

\bibitem{Burghardt1999}
I.~Burghardt, H.-D. Meyer, and L.~S. Cederbaum.
\newblock {Approaches to the approximate treatment of complex molecular systems
  by the multiconfiguration time-dependent Hartree method}.
\newblock {\em The Journal of Chemical Physics}, 111(7):2927, 1999.

\url {http://dx.doi.org/10.1063/1.479574}

\bibitem{Worth2003a}
G.~A. Worth and I.~Burghardt.
\newblock {Full quantum mechanical molecular dynamics using Gaussian
  wavepackets}.
\newblock {\em Chemical Physics Letters}, 368:502, 2003.

\url {http://dx.doi.org/10.1016/S0009-2614(02)01920-6}

\bibitem{Shalashilin2008}
D.~V. Shalashilin and I.~Burghardt.
\newblock {Gaussian-based techniques for quantum propagation from the
  time-dependent variational principle: Formulation in terms of trajectories of
  coupled classical and quantum variables}.
\newblock {\em The Journal of Chemical Physics}, 129(8):084104, 2008.

\url {http://dx.doi.org/10.1063/1.2969101}

\bibitem{Note08}
For coherent states other than canonical, $|\braket{z}{z'}|$ is no longer a
Gaussian distribution, but is still localized in phase space.

\bibitem{Note09}
Latin letters $j,k,l$ will be used for labeling basis-set elements and, for
convenience, we henceforth abbreviate basis-set summations by omitting their
ranges.

\bibitem{Kay1989}
K.~G. Kay.
\newblock {The matrix singularity problem in the time-dependent variational
  method}.
\newblock {\em Chemical Physics}, 137:165, 1989.

\url {http://dx.doi.org/10.1016/0301-0104(89)87102-2}

\bibitem{Habershon2012a}
S.~Habershon.
\newblock {Linear dependence and energy conservation in Gaussian wavepacket
  basis sets}.
\newblock {\em The Journal of Chemical Physics}, 136(1):014109, 2012.

\url {http://dx.doi.org/10.1063/1.3671978}

\bibitem{Milburn1997}
G.~J. Milburn, J.~Corney, E.~M. Wright, and D.~F. Walls.
\newblock {Quantum dynamics of an atomic Bose-Einstein condensate in a
  double-well potential}.
\newblock {\em Physical Review A}, 55(6):4318, 1997.

\url {http://dx.doi.org/10.1103/PhysRevA.55.4318}

\bibitem{Viscondi2009}
T.~F. Viscondi, K.~Furuya, and M.~C. de~Oliveira.
\newblock {Generalized purity and quantum phase transition for Bose-Einstein
  condensates in a symmetric double well}.
\newblock {\em Physical Review A}, 80:013610, 2009.

\url {http://dx.doi.org/10.1103/PhysRevA.80.013610}

\bibitem{Viscondi2009a}
T.~F. Viscondi, K.~Furuya, and M.~C. de~Oliveira.
\newblock Coherent state approach to the cross-collisional effects in the
  population dynamics of a two-mode {Bose-Einstein} condensate.
\newblock {\em Annals of Physics}, 324(9):1837, 2009.

\url {http://dx.doi.org/10.1016/j.aop.2009.05.008}

\bibitem{Sakmann2009}
K.~Sakmann, A.~I. Streltsov, O.~E. Alon, and L.~S. Cederbaum.
\newblock Exact quantum dynamics of a bosonic josephson junction.
\newblock {\em Phys. Rev. Lett.}, 103:220601, 2009.

\url {http://dx.doi.org/10.1103/PhysRevLett.103.220601}

\bibitem{Schwinger}
J.~Schwinger.
\newblock On angular momentum.
\newblock In L.~C. Biedenharn and H.~Van~Dam, editors, {\em Quantum Theory of
  Angular Momentum: A Collection of Reprints and Original Papers}, page 229.
  Academic Press, London, 1965.

\url {http://dx.doi.org/10.2172/4389568}

\bibitem{Note10}
Throughout this section units are such that $\hbar=1$.

\bibitem{Note11}
To compute the Hamiltonian \eqref{eqn:2well-hamiltonian-3}, one could either use
decomposition \eqref{eqn:su2-cs-1} to evaluate the required matrix elements of
angular momentum operators or employ $\mathrm{SU}(2)$ generating functions.
\cite{Zhang1990a}

\bibitem{Note12}
This can be verified by calculating the standard deviations of an appropriate
quantum phase-space distribution with respect to the angular coordinates
$\theta$ and $\phi$.

\bibitem{Yaffe82}
L.~G. Yaffe.
\newblock Large N limits as classical mechanics.
\newblock {\em Reviews of Modern Physics}, 54(2):407, 1982.

\url {http://dx.doi.org/10.1103/RevModPhys.54.407}

\bibitem{Viscondi2010}
T.~F. Viscondi, K.~Furuya, and M.~C. de~Oliveira.
\newblock {Phase transition, entanglement and squeezing in a triple-well
  condensate}.
\newblock {\em EPL (Europhysics Letters)}, 90:10014, 2010.

\url {http://stacks.iop.org/0295-5075/90/i=1/a=10014}

\bibitem{Viscondi11a}
T.~F. Viscondi and K.~Furuya.
\newblock Dynamics of a {Bose-Einstein} condensate in a symmetric triple-well
  trap.
\newblock {\em Journal of Physics A: Mathematical and Theoretical}, 44:175301,
  2011.

\url {http://stacks.iop.org/1751-8121/44/i=17/a=175301}

\bibitem{Note13}
We note that in the triple-well model, the energy difference between the
ground state and doubly degenerate excited states of the one-body Hamiltonian is
$3\hbar\Omega$ -- within the three-mode approximation these stationary states
span the same eigenspace as the local modes associated with operators $a_1$,
$a_2$ and $a_3$. \cite{Viscondi11a} Also, in Eq.
\eqref{eqn:3-well-hamiltonian-1}, we have made an additional simplification by
excluding cross-collision terms, which arise from the interaction between bosons
in different wells.

\bibitem{Note14}
This means that the smaller basis sets are embedded in the larger ones.

\bibitem{Note15}
No significant improvement of the results was observed for larger basis sets
constructed with the same sampling parameters.

\bibitem{Note16}
As can be seen from Eq.~\eqref{eqn:sun-cs-1}, for the particular case of $N =
1$, the bosonic $\mathrm{SU}(n)$ coherent state reduces to $\ketn{z} =
\ket{\phi_n} +\sum^{n-1}_{\alpha=1} z_\alpha \ket{\phi_\alpha}$, where the
occupation number representation has been replaced by `first-quantized'
notation, with $\ket{\phi_\alpha}$ being the single-particle orbital associated
with the $\alpha$-th mode. This is nothing but a standard decomposition of a
quantum state in a finite basis of size $n$, only written without redundant
parameters.

\bibitem{Shalashilin2010}
D.~V. Shalashilin.
\newblock {Nonadiabatic dynamics with the help of multiconfigurational
  Ehrenfest method: Improved theory and fully quantum 24D simulation of
  pyrazine}.
\newblock {\em The Journal of Chemical Physics}, 132(24):244111, 2010.

\url {http://dx.doi.org/10.1063/1.3442747}

\bibitem{Saita2012}
K.~Saita and D.~V. Shalashilin.
\newblock {On-the-fly ab initio molecular dynamics with multiconfigurational
  Ehrenfest method}.
\newblock {\em The Journal of Chemical Physics}, 137(22):22A506, 2012.

\url {http://dx.doi.org/10.1063/1.4734313}

\bibitem{GSLmanual}
M.~Galassi, J.~Davies, J.~Theiler, B.~Gough, G.~Jungman, M.~Booth, and
  F.~Rossi.
\newblock {\em {Gnu Scientific Library: Reference Manual}}.
\newblock Network Theory Ltd., 3rd edition, 2003.

\url {http://www.gnu.org/software/gsl/}

\bibitem{Note17}
The overlap matrix is Hermitian and positive-definite, meaning that its
eigenvalues are real and positive, though numerical diagonalization may produce
null or very small negative eigenvalues. Alternatively, one could employ a
singular value decomposition and carry on the sampling procedure using the
singular values rather than the eigenvalues.

\end{thebibliography}
\end{document}